\documentclass[preprints,review,accept,moreauthors,pdftex,10pt,a4paper]{mdpi} 

\firstpage{1} 
\pubvolume{xx}
\issuenum{8}
\articlenumber{607}
\pubyear{2022}
\copyrightyear{2022}
\externaleditor{Academic Editor: Ioana Dutan and Nicholas R. MacDonald}
\history{Received: date; Accepted: 15 November 2022; Published: date}

\pdfoutput=1

\Title{Active Galactic Nuclei as potential Sources of Ultra-High Energy Cosmic Rays}
 
\Author{Frank M. Rieger $^{1,2}$\orcidA{}}

\address{%
$^{1}$ \quad Institute for Theoretical Physics (ITP), Heidelberg University, Philosophenweg 12, 
69120 Heidelberg, Germany; f.rieger@uni-heidelberg.de\\
$^{2}$ \quad Max-Planck-Institut für Kernphysik (MPIK), P.O. Box 103980, 69029 Heidelberg, 
Germany}

\corres{Correspondence: f.rieger@uni-heidelberg.de}

\abstract{Active Galactic Nuclei (AGNs) and their relativistic jets belong to the most promising 
class of ultra-high-energy cosmic ray (UHECR) accelerators. This compact review summarises 
basic experimental findings by recent instruments, and discusses possible interpretations and 
astrophysical constraints on source energetics. Particular attention is given to potential sites 
and mechanisms of UHECR acceleration in AGNs, including gap-type particle acceleration 
close to the black hole, as well as first-order Fermi acceleration at trans-relativistic shocks and 
stochastic shear particle acceleration in large-scale jets. It is argued that the last two represent 
the most promising mechanisms given our current understanding, and that nearby FR~I type
radio galaxies provide a suitable environment for UHECR acceleration.}

\keyword{Ultra high energy cosmic rays; Particle acceleration; Radio Galaxies, Relativistic 
jets.} 

\newcommand{\gppr}{\stackrel{>}{\scriptstyle \sim}}
\newcommand{\lppr}{\stackrel{<}{\scriptstyle \sim}}
\begin{document}

\section{Introduction}
The energy spectrum of cosmic rays runs over more than ten orders of magnitudes, from 
GeV energies to $\sim 10^{20}$ eV. While supernova remnants are believed to be the most 
probable sources of cosmic rays at lower energies (i.e., up to the 'knee' at $\sim 3\times 
10^{15}$ eV)\cite{Bell2013,Blasi2013}, the origin of ultra-high-energy cosmic rays (UHECRs, 
$E\geq 10^{18}$ eV = 1 EeV) is much less understood. While thought to be of extragalactic 
origin \cite{Aab2017}, the real astrophysical sources are still to be deciphered. Possible 
candidate sources include Active Galactic Nuclei (AGNs) and their relativistic jets, starburst 
galaxies (SBGs) and gamma-ray bursts, as well as accretion shocks around clusters of 
galaxies, see e.g., ref.~\cite{Kotera2011}. 
The present review will focus on the relevance of AGNs, and highlight some of the recent 
developments in the field, including a compact discussion of particle acceleration in the 
relativistic jets of AGNs. As such it draws on an earlier review \cite{Rieger2019b}, 
extending and updating its perspectives.

Jetted AGNs \cite{Padovani2017} have for long be considered as promising UHECR emitters. 
As the most powerful, persistent sources in the Universe, they are in principle capable to satisfy 
the physics constraints (e.g., size, energetics) on putative UHECR emitters, see e.g., 
ref.~\cite{Hillas1984}. 
Since UHECRs lose energy during propagation by inelastic collisions with CMB photons, possible 
AGN sources have to be relatively nearby (i.e., within a few hundred Mpc)\footnote{This so-called 
GZK horizon is longest ($\sim 250$ Mpc) for 100 EeV protons and heavy nuclei (e.g., iron), but 
can be much shorter for intermediate masses (e.g., $\sim 5$ Mpc for helium and $\sim 100$ Mpc 
for silicon).} 
in order to be able to account for the highest cosmic-ray energies observed. In recent years, 
particular attention has thus been given to the nearest AGNs, i.e., the radio galaxies Centaurus A
(Cen~A), M87 and Fornax A. As shown below, these sources are indeed particularly interesting 
given current experimental results and theoretical considerations.

From an experimental point of view, major progress in UHECR research has been achieved over the 
last decade by the Pierre Auger (PA) and the Telescope Array (TA) collaborations, see e.g., 
refs.~\cite{Rubtsov2022,Novotny2022} for recent experimental reports. Both collaborations run hybrid, 
large-scale instruments that are fully operational since 2008. The larger, PA observatory consists of 1660 
surface (water tank) detectors and four fluorescence detector stations, and is located in the southern 
hemisphere (Mendoza, Argentina). The original TA construction, on the other hand, consists of 507 surface 
(scintillation counter) detectors and three fluorescence detector stations, and is located in the northern 
hemisphere (Utah, USA). Both instruments are "hybrid" in nature: Their surface detectors (SDs) measure 
air shower particles on the ground, and are sensitive to its electromagnetic, muonic and hadronic 
components, while the air fluorescence detectors (FDs) observe the longitudinal development of air 
showers in the atmosphere by the light emitted during their passage. The FDs provide a direct 
calorimetric measurement of the shower energy, and can thus be used to calibrate the energy scale 
of the SD events.

In the following, selected experimental results are briefly summarized. For comprehensive overviews 
and discussion of recent results the reader is also referred to refs.~\cite{Aloisio2017,Kachelriess2019,
Anchordoqui2019,Kachelriess2022,Coleman2022}.

\section{Basic Results of current UHECR Experiments}
\subsection{Energy Spectrum of UHECRs}\label{sec_spectrum}
\begin{figure}[htbp]
\begin{center}
\includegraphics[width = 0.49 \textwidth]{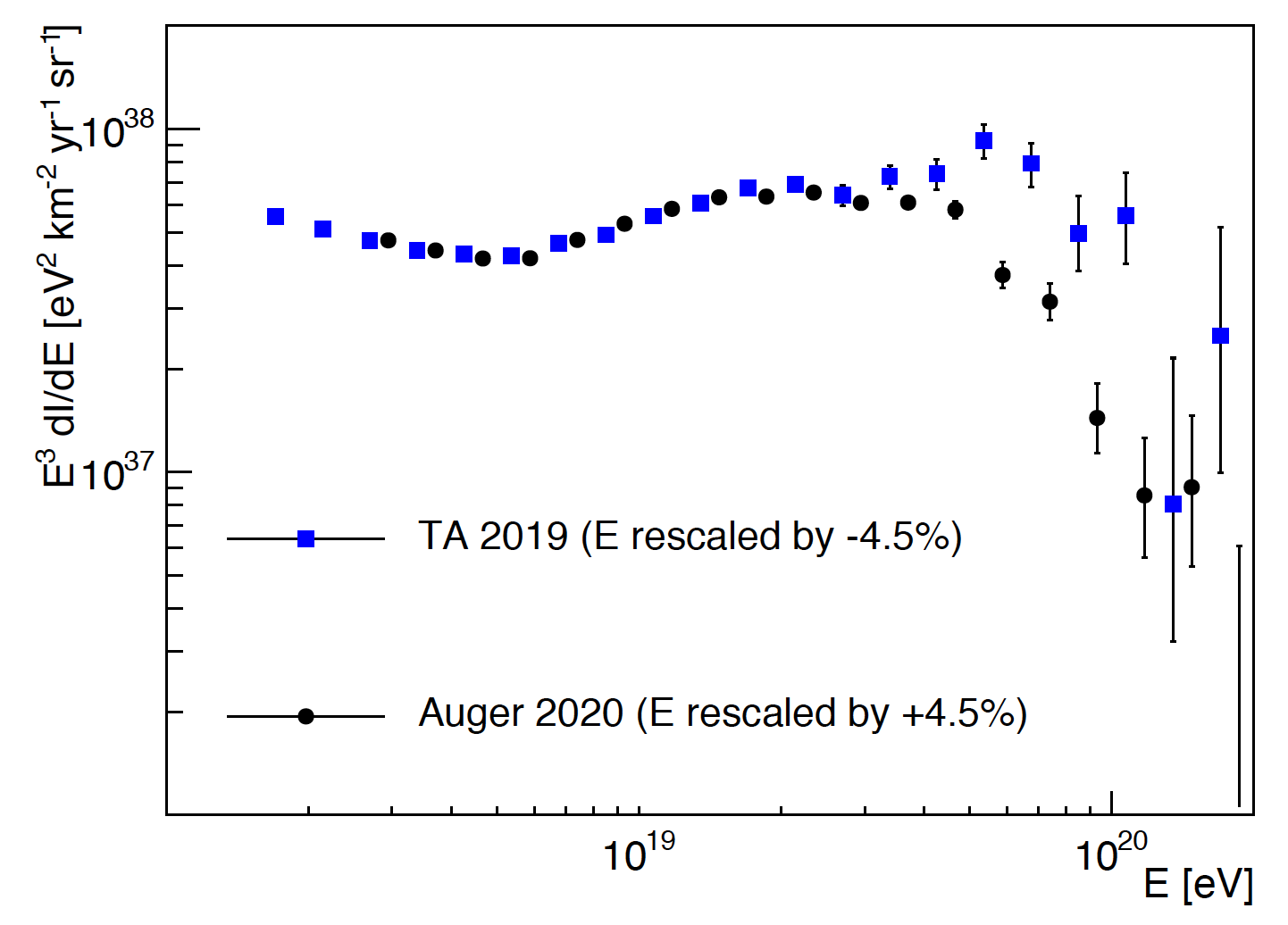}
\includegraphics[width = 0.49 \textwidth]{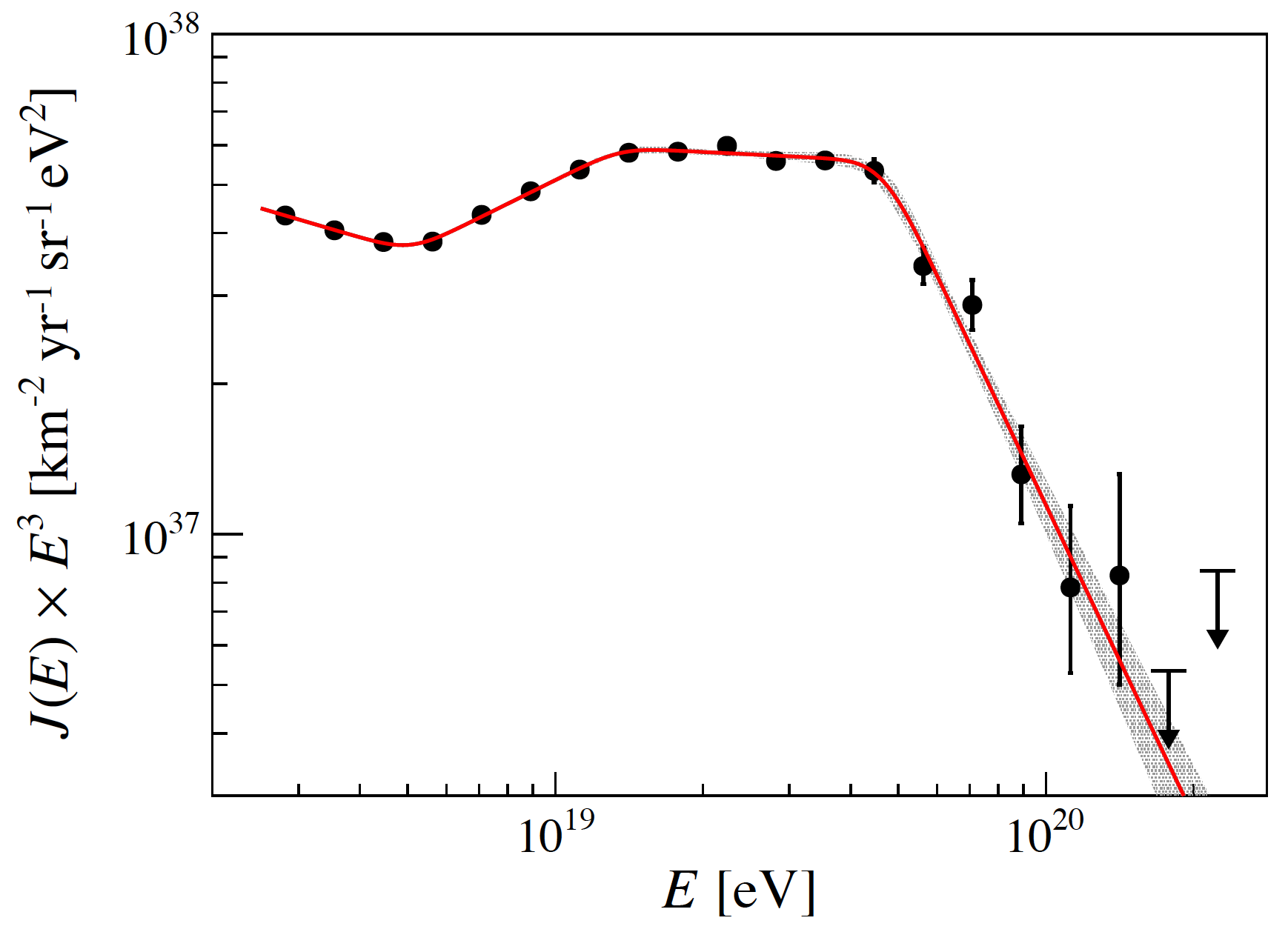}
\caption{{\it Left:} Energy spectrum of UHECRs as measured by the PA (Auger) and TA surface detectors, 
with each spectrum being adjusted for uncertainties in absolute energy scale by the same amount. Good 
agreement is achieved up to $E\sim 4\times 10^{19}$ eV. 
{\it Right:} Evidence from PA data for a new spectral feature ("instep") between $1.5\times 10^{19}$ and 
$4.5\times 10^{19}$, also supported by TA data. The observed spectrum can be well fitted by a series of 
four power-laws. From ref.~\cite{Tsunesada2022}.}
\label{spectrum1}
\end{center}
\end{figure}

The energy spectrum of UHECRs extends up to $\sim 10^{20}$ eV, with the highest-energy event reported 
having an energy of about $3 \times 10^{20}$ eV \cite{bird1995}. A comparison of current PA and TA 
measurements shows that up to $E\sim 4\times 10^{19}$ eV a good spectral agreement between experiments, 
including the position of the "ankle" at $\simeq  5\times 10^{18}$ eV, is achieved once the uncertainty 
($\sim10\%$) in absolute energy scale is taken into account, see e.g., \cite{Tsunesada2022} and 
Figure~\ref{spectrum1}. 
Whether possible differences towards higher energies are simply caused by different instrumental 
systematics (including energy-dependent effects), and/or some real source differences (e.g., as the 
particle mean free path decreases, differences in the 'large-scale' structure may no longer be negligible) 
still remains to be studied. 
In general, the overall UHECR energy spectrum reveals a pronounced steepening above $E\sim 5\times 
10^{19}$ eV, that is compatible with a Greisen-Zatsepin-Kuzmin (GZK) cut-off \cite{Greisen1966,Zatsepin1966}, 
but could also be related to the maximum particle acceleration efficiency in UHECR emitting, astrophysical  
sources. A closer look now provides evidence for an additional spectral feature (referred to as "instep") 
between $1.5\times 10^{19}$ eV and $4.5\times 10^{19}$ eV (with power-law index $p\simeq 3$, where 
$J(E)\propto E^{-p}$). The spectrum becomes harder after the ankle and before $1.5\times 10^{19}$ eV 
(index $p$ changing from $p\simeq 3.3 \rightarrow 2.5$), and significantly steepens above $4.5\times 
10^{19}$ eV (index $p$ changing from $p \simeq 3 \rightarrow 5$), cf. Figure~\ref{spectrum1}. 
This could be interpreted as providing some indications for multiple astrophysical source contributions to the 
overall UHECR spectrum, such as, e.g., two extragalactic components, a soft lower-energy and a fiducial, 
hard (index $p<0$) higher-energy one \cite{Guido2022}.
For reference, the observed UHECR spectrum roughly translates into an UHECR luminosity density 
(around $10^{19.5}$ eV) of $l_{\rm UHECR} \sim 6\times 10^{43}$ erg/(Mpc$^3$ yr) \cite{Murase2019}.

\subsection{Composition of UHECRs}\label{sec_composition}
\begin{figure}[htbp]
\begin{center}
\includegraphics[width = 0.99 \textwidth]{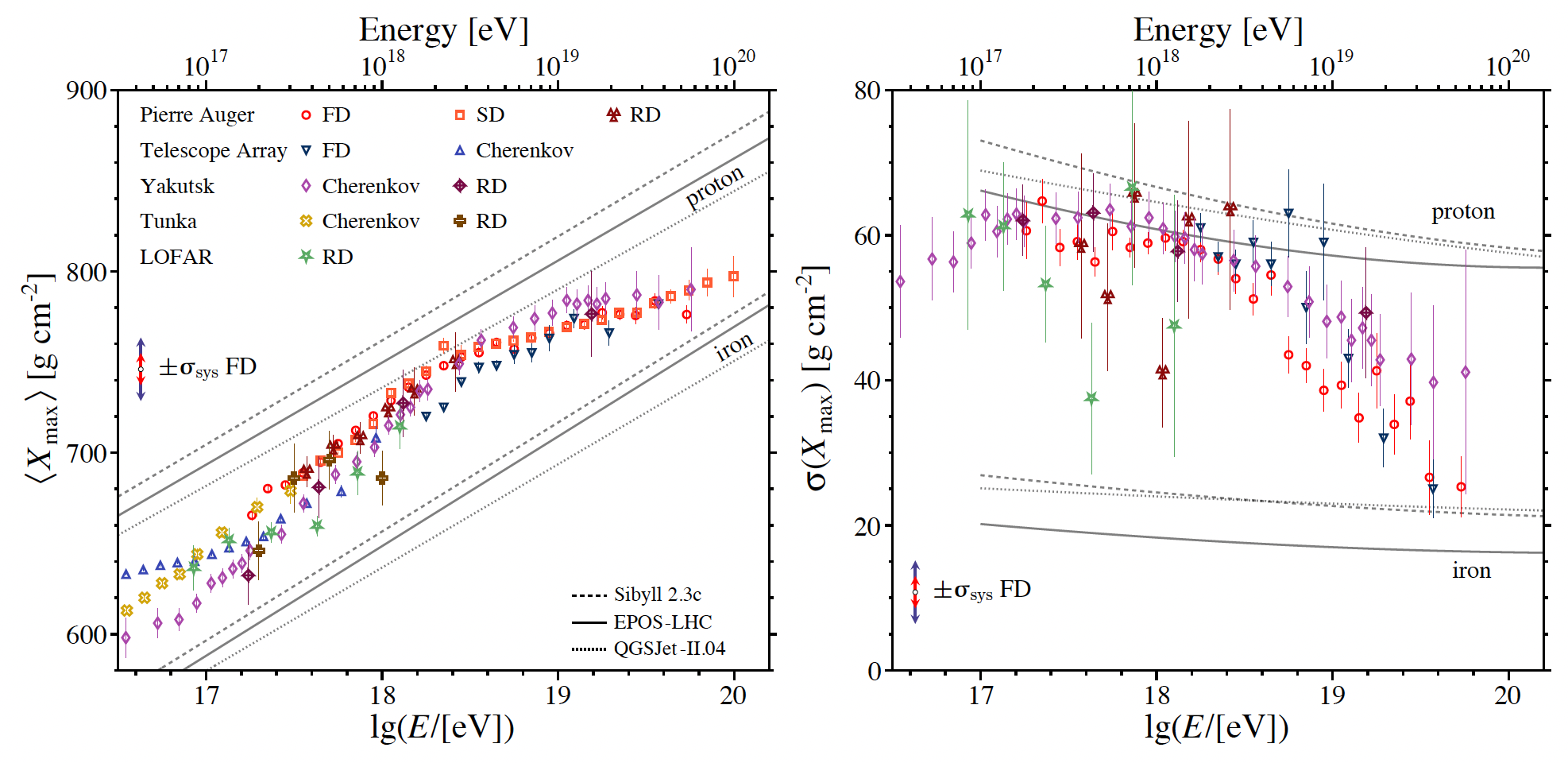}
\caption{Measurements of the mean (left) and its fluctuations (right) of the distribution of shower 
maximum $X_{\rm max}$ as function of energy, along with predictions for proton and iron nuclei 
based on hadronic interaction models. There is an apparent trend for the composition to become 
heavier towards highest energies. Systematic uncertainties of the fluorescence (FD) measurements 
of TA (Telescope Array) and PA (Auger) are indicated as blue and red arrows, 
respectively. From ref.~\cite{Coleman2022}.}
\label{composition}
\end{center}
\end{figure}

Upon entering the atmosphere, cosmic-ray particles initiate hadronic showers by interaction with atmospheric 
nuclei, cf. ref.~\cite{Engel2011} for review. The number of secondary shower particles increases up to a depth 
$X_{\rm max}$ (measured in g/cm$^2$), beyond which the available energy becomes insufficient to create 
further particles. In general, the depth of the shower maximum $X_{\rm max}$ is sensitive to the primary mass 
$A$ and initial energy $E_0$, approximately as $X_{\rm max} \propto \ln[E_0/A]$, and hence light particles 
penetrate deeper in the atmosphere and exhibit a steeper lateral distribution compared to heavy nuclei. 
The UHECR mass composition can thus in principle be inferred from measurements of the mean, $\langle 
X_{\rm max} \rangle$, and its dispersion (RMS), $\sigma(X_{\rm max})$. While high-energy hadronic 
interaction models play an important role in this and a coherent explanation of all data still remains 
challenging \cite{Ostapchenko2019}, the currently favoured interpretation points to a primary mass 
composition that is mixed and changes with energy. In particular, a comparison of UHECR data suggests 
that the composition becomes lighter between $10^{17.2}$ eV and $10^{18.2}$ eV, compatible with a 
transition from galactic to extragalactic cosmic rays \cite{Coleman2022}.
Above $E\simeq 10^{18.2}  \sim 2\times 10^{18}$ eV the composition seems to become heavier again, as 
evident from the measurements of both, the mean $X_{\max}$ and its fluctuations $\sigma(X_{max})$ (see 
Figure~\ref{composition}). There is evidence that protons are gradually replaced by helium, helium by 
nitrogen etc, possibly continuing up to iron (e.g., ref.~\cite{Alves_Batista2019}).
Note that within current systematic and statistical uncertainties, the TA and PA results appear compatible 
with each other. As shown below, the requirements on astrophysical accelerators become much less 
demanding if the composition becomes indeed heavier towards the highest energies.

\subsection{Anisotropies in the arrival direction of UHECRs}\label{sec_anisotropy}
While UHECRs are deflected in magnetic fields, significant anisotropies in their arrival directions could offer
important clues as to their astrophysical origin. On large angular scales ($> 40^{\circ}$), for example, PA has 
reported strong evidence ($\geq 6 \sigma$) for a dipole anisotropy ($\sim 115^{\circ}$ away from the Galactic 
Center) above $8\times 10^{18}$ eV with an amplitude of $\simeq 7\%$, supporting an extragalactic origin of 
UHECRs \cite{Aab2017,Almeida2022}. The $4\sigma$ evidence for a growth of the amplitude with energy
further supports this claim \cite{PAO2018_aniso}.
In principle, searches for possible anisotropies on intermediate scales (a few tens of degrees) could offer further, 
improved insights. Typically however, these anisotropy results are somewhat less significant and need thus be 
viewed with some caution. Nevertheless, based on an updated analysis, TA \cite{Kim2022,Tkachev2022} has 
reported evidence for a "hot spot" in the UHECR events above $5.7\times 10^{19}$ eV (smeared out on circles of 
$25^{\circ}$) at a level of $\sim 5 \sigma$ (local significance) and $3.2 \sigma$ (global significance), respectively 
(cf. also \cite{Matthews2017}). In equatorial coordinates, the TA hot spot is centered around R.A. $\sim144^{\circ}$ 
and dec $\sim41^{\circ}$), approximately in the direction of the starburst galaxy M82 ($d\sim 3.8$ Mpc) and the 
Ursa Major/Virgo supercluster ($d \sim18$ Mpc). In addition, at lower energies $E\geq 2.5\times 10^{19}$ a 
"warm spot" ($\sim 4 \sigma$ local significance) around R.A. $\sim 18^{\circ}$ and dec $\sim35^{\circ}$, 
approximately in the direction of the Perseus-Pisces supercluster ($d\sim 70$ Mpc), or Andromeda 
\cite{Zirakashvili2022}, has been reported. On the other hand, based on a constrained search PA has also 
reported an UHECR excess ("warm spot") at a $\sim 4\sigma$ level above $E>4\times 10^{19}$ eV on an 
angular scale of $27^{\circ}$ around the Centaurus region \cite{Biteau2022,PAO2022}, containing amongst 
other objects the radio galaxy Cen~A and the starburst galaxies M83 and NGC 4945, see also 
Figure~\ref{correlations}.

\section{On possible Correlations of UHECRs with Matter and Known Sources}
The anisotropy findings have motivated a variety of studies searching for possible correlations with 
matter and/or AGNs and starburst galaxies catalogs. Examples include: 

\subsection{On Correlations with Catalogs:}
\begin{figure}[htbp]
\begin{center}
\includegraphics[width = 0.85 \textwidth]{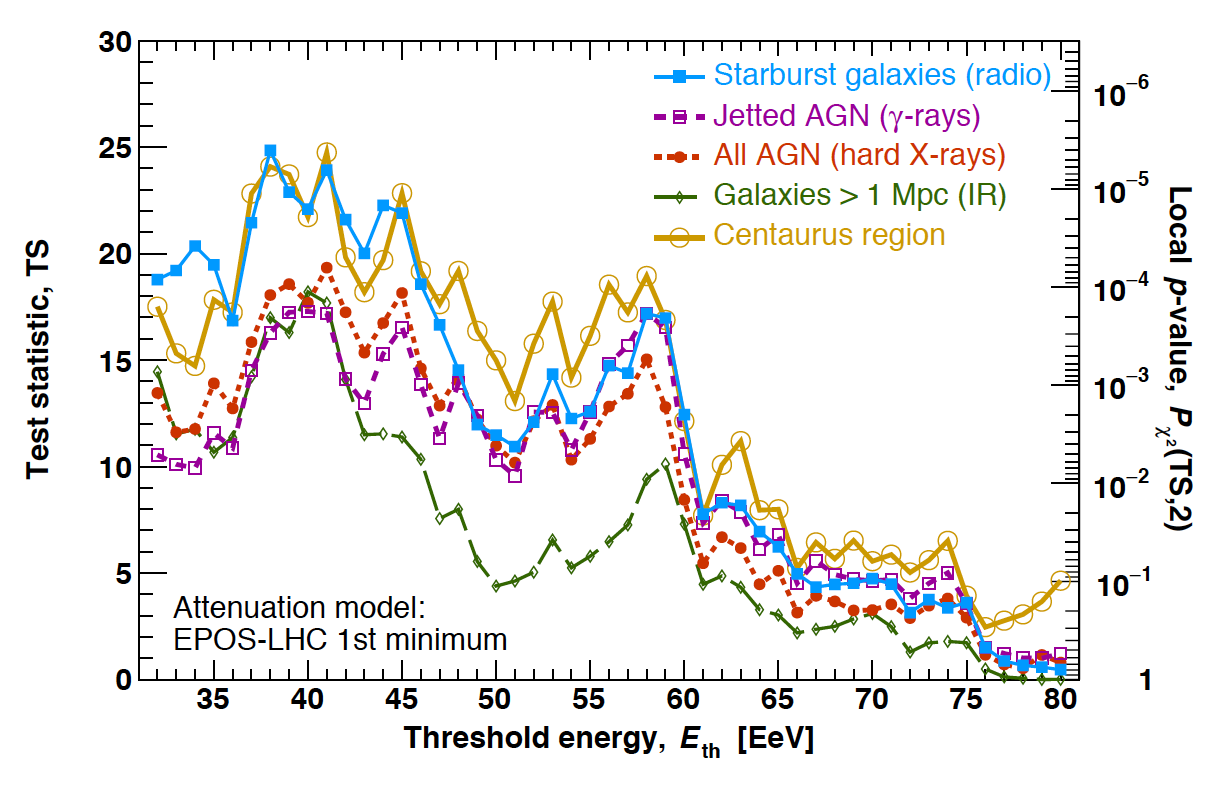}
\caption{Significance (test statistic) of the search for UHECR anisotropies as a function of threshold 
energy (in EeV) against four, complementary catalogs, including jetted AGNs and starburst galaxies 
(SBGs), along with the results for the Centaurus region. In general, the AGN catalogs are dominated 
by a contribution from Cen~A, while the SBGs M83 and NGC 4945 in the Centaurus region feature 
prominently in the SBG model. From ref.~\cite{Biteau2022}.}
\label{correlations}
\end{center}
\end{figure}

In the most recent PA analysis, correlation studies of the arrival direction of PA events above $32$ EeV 
($\sim 2600$ events) have been performed with four different catalogs, including a catalog of jetted AGNs 
and a catalog of starburst galaxies (SBGs) \cite{Biteau2022,PAO2022}. 
Interestingly, a SBG model with search radius $\sim 25^{\circ}$ that attributes $9\%$ of the UHECR ($>38$ 
EeV) events to a catalog of 44 nearby SBGs (including NGC 4945, NGC 253, M83, NGC 1068), and the 
remaining to an isotropic background, appears favoured at $4\sigma$ over the hypothesis of isotropy.
In comparison, a jetted AGN model with search radius of $\sim 23^{\circ}$ that attributes $\sim 6\%$ of 
the UHECR flux to a catalog of 26 nearby, $\gamma$-bright AGNs (including Cen A, M87, Mkn 421, 
Mkn 501 and Fornax A) is only preferred at a $3\sigma$ level, cf. Figure~\ref{correlations}. 
However, although the significance is different, these findings do not (yet) provide statistically compelling 
evidence for a catalog preference. 
From an astrophysical physical point of view, the interpretation is further complicated by the fact that 
(i) the UHECR luminosities of individual AGNs or SBGs is not known and thus some ad-hoc proxies 
(e.g., average Fermi-LAT fluxes for AGNs) have to be employed and (ii) a realistic treatment of 
deflection in magnetic fields is not incorporated as variations from model to model remain too 
large to be included in the analysis. Note that even if a significant statistical association of 
UHECR events with SBGs (e.g., the TA hot spot with M82) were to be established, this would 
physically not necessarily be conclusive, as it might also be related to an echo of the past, i.e., 
UHECRs that were accelerated by radio galaxies and subsequently being scattered by SBGs 
(see 3.4. below) \cite{Bell2022}.

\subsection{On Correlation with Matter:}
Following a different approach, Ding et al.~\cite{Ding2021} have recently studied the possible imprint
of the (local) large-scale structure on the UHECR sky distribution. Assuming that the UHECR source 
density is proportional to the Cosmicflows-2 matter density field \cite{Hoffman2018} and incorporating 
the PA-inferred chemical composition of UHECRs, they showed that the magnitude and the 
energy-dependence and, to some extent, also the direction of  the PA dipole anisotropy above $E> 
8\times 10^{18}$ eV could in principle be reproduced in a scenario with many low power sources 
(following the matter distribution) instead of a few high-power ones. Their findings verify that the 
observed anisotropy is incompatible with a pure proton composition. Typically, in such frameworks 
a good reconstruction of the dipole direction is quite challenging to achieve \cite{Allard2021}. The 
considered approach is advanced in that it includes the relevant energy losses and magnetic 
deflection (essentially in the Galactic magnetic field) via multi-particle tracking as well as fits to the 
(PA) mass composition. Possible caveats concern the use of a highly simplified source model (all 
being standard candles in terms of luminosity, composition and spectral index), the fact that 
prominent individual sources (e.g., Cen~A, cf. above) are not included, and the assumption of a 
homogeneous turbulent (non-structured) extragalactic magnetic field. 

\subsection{On Correlation with Nearby Radio Galaxies:}
\begin{figure}[htbp]
\begin{center}
\includegraphics[width = 0.98 \textwidth]{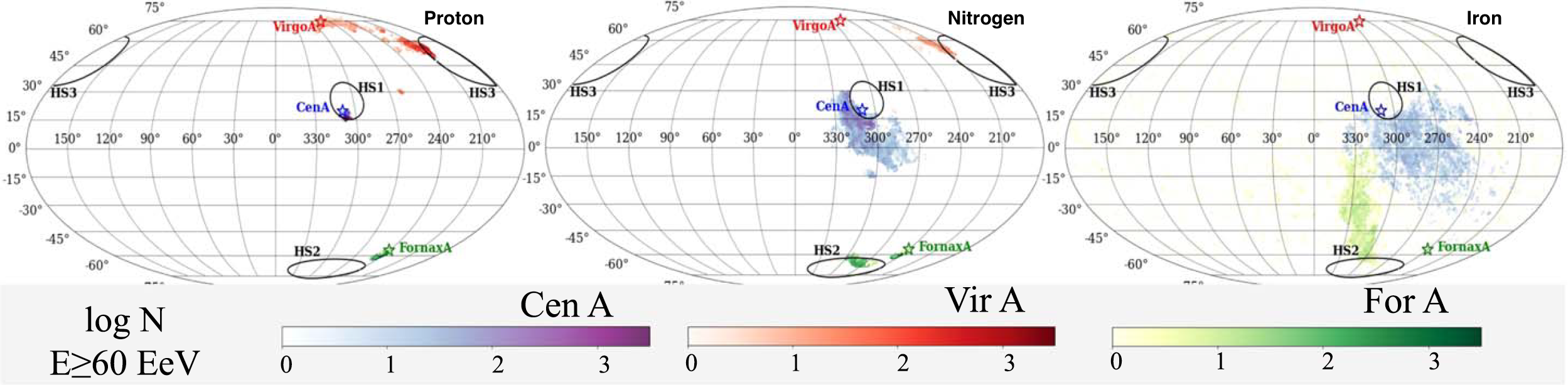}
\caption{Sky maps (galactic coordinates) of the arrival direction of simulated UHECR events above 
60 EeV for a strong and highly structured extragalactic magnetic field (Prim), along with the locations 
of three excess (HS: 'hot spots') regions inferred from PA and TA maps (black solid circles). From 
left to right for proton, nitrogen and iron. The contributions from M87 (Virgo A), Cen A and Fornax A 
are shown in red, blue and green, respectively. 
Accordingly, light nuclei from Cen A may contribute to HS1, intermediate nuclei from Fornax A 
to HS2 and light to intermediate nuclei from M87 to HS3. From ref.~\cite{Oliveira2022}.}
\label{UHECR_RG}
\end{center}
\end{figure}

Using structured extragalactic magnetic field (EGMF) models with voids and filaments as inferred 
from MHD (ENZO) simulations, and focusing on nearby radio galaxies, de Oliveira \& de Souza 
\cite{Oliveira2022} have recently shown that the EGMF structure can dominate the arrival of
UHECRs from nearby sources. In particular, while the EGMF choice does not strongly affect the 
UHECR flux from Cen A, Earth is found to be in a flux-suppressed position to receive UHECRs 
from the direction of M87, while favoured for Fornax A. With only these three sources, the PA 
dipole direction cannot be explained at energies below $32$ EeV, but might be above these energies
if the sources inject heavy nuclei. Moreover, Cen~A, M87 and Fornax A are found to contribute to the 
excess emission associated with the detected UHECR hot spots (HS). In particular, Cen A may 
contribute to HS1 (by light nuclei), Fornax A to HS2 (by intermediate nuclei) and M87 to HS3 
(by light to intermediate nuclei), cf. Figure~\ref{UHECR_RG}.
The considered model does not aim to fit the mass composition, but includes the relevant energy 
losses and a (CRPropa3) treatment of magnetic deflection in the Galactic and the local EGMF, as 
well as source specific rigidity cut-offs. The results fit into a framework where a few nearby radio 
galaxies are considered to dominate the UHECR flux above the ankle, cf. also \cite{Eichmann2022}.

\subsection{An UHECR echo from the past?} 
As common in astrophysics, finding some correlation does not straightforwardly establish causality, 
in our case the true astrophysical sites of UHECR acceleration. This introduce a general caveat 
for such kind of approaches. As has been argued recently, for example, a putative correlation of 
UHECRs with starburst galaxies might be the echo of UHECRs originally produced in Cen A 
\cite{Bell2022}, see Figure~\ref{CenA_echo} below. In this scenario, some of the UHECRs accelerated 
in Cen A, are considered to propagate directly to Earth, while others arrive later via M82 and/or M81, 
by being scattered and diffusively reflected in their halos. This would then imply that the present flux 
of UHECRs should rather be attributed to a leakage from the lobes of Cen A, with these lobes 
representing a reservoir of UHECRs accelerated during earlier (likely more powerful) activity 
episodes.
\begin{figure}[h]
\begin{center}
\includegraphics[width=0.60 \textwidth]{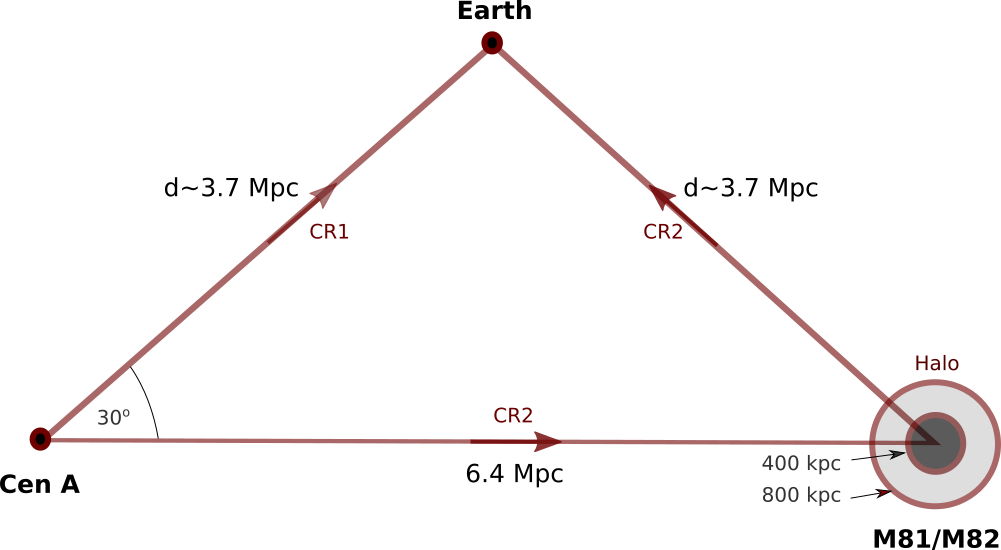}
\caption{Sketch of the echo model in which UHECRs, originally being accelerated in Cen A, 
are scattered and reflected in the spherical halo of the starburst galaxy M82 (with radius up 
to 800 kpc), introducing an apparent UHECR - starburst galaxy correlation. In this scenario, 
UHECRs arriving at Earth via M82 (CR2) are delayed by approximately $20$ Myr relative 
to those arriving by direct propagation (CR1) from Cen A. 
After ref.~\cite{Bell2022}.}\label{CenA_echo}
\end{center}
\end{figure}

\section{Radio Galaxies as Prime UHECR Candidate Sources}
\begin{figure}[h]
\begin{center}
\includegraphics[width=0.90 \textwidth]{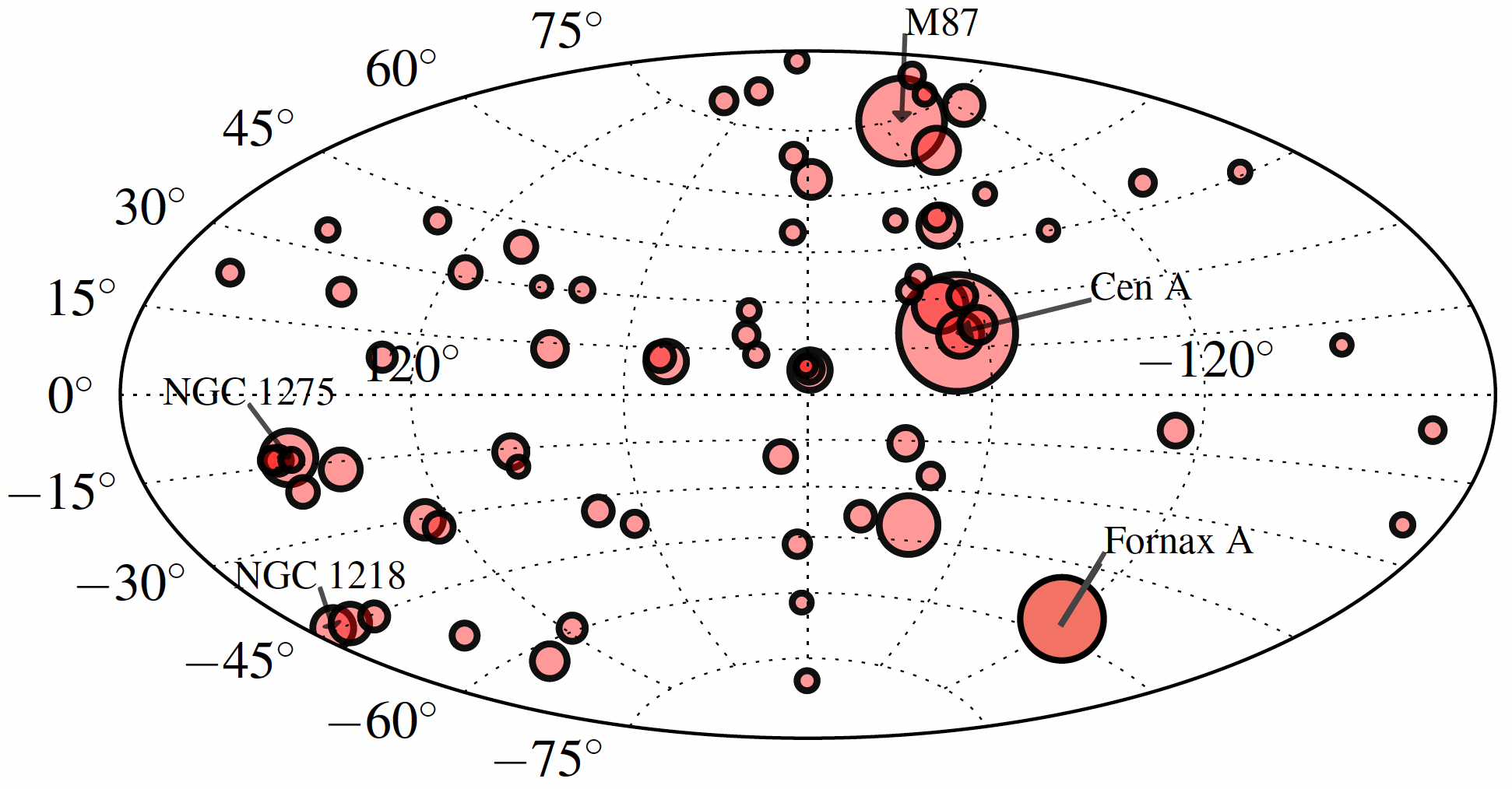}
\caption{Map in Galactic coordinates of the 74 radio galaxies within $\sim 130$ Mpc, along with the
locations of some prominent (FR~I) sources. The circular areas are proportional to the $\sim 1$ 
GHz-radio flux of the source. Adapted (with Fornax A adjusted) based on ref.~\cite{VanVelzen2012}.}\label{local_RG}
\end{center}
\end{figure}

\begin{figure}[h]
\vspace*{-0.4cm}
\begin{center}
\includegraphics[width=0.98 \textwidth]{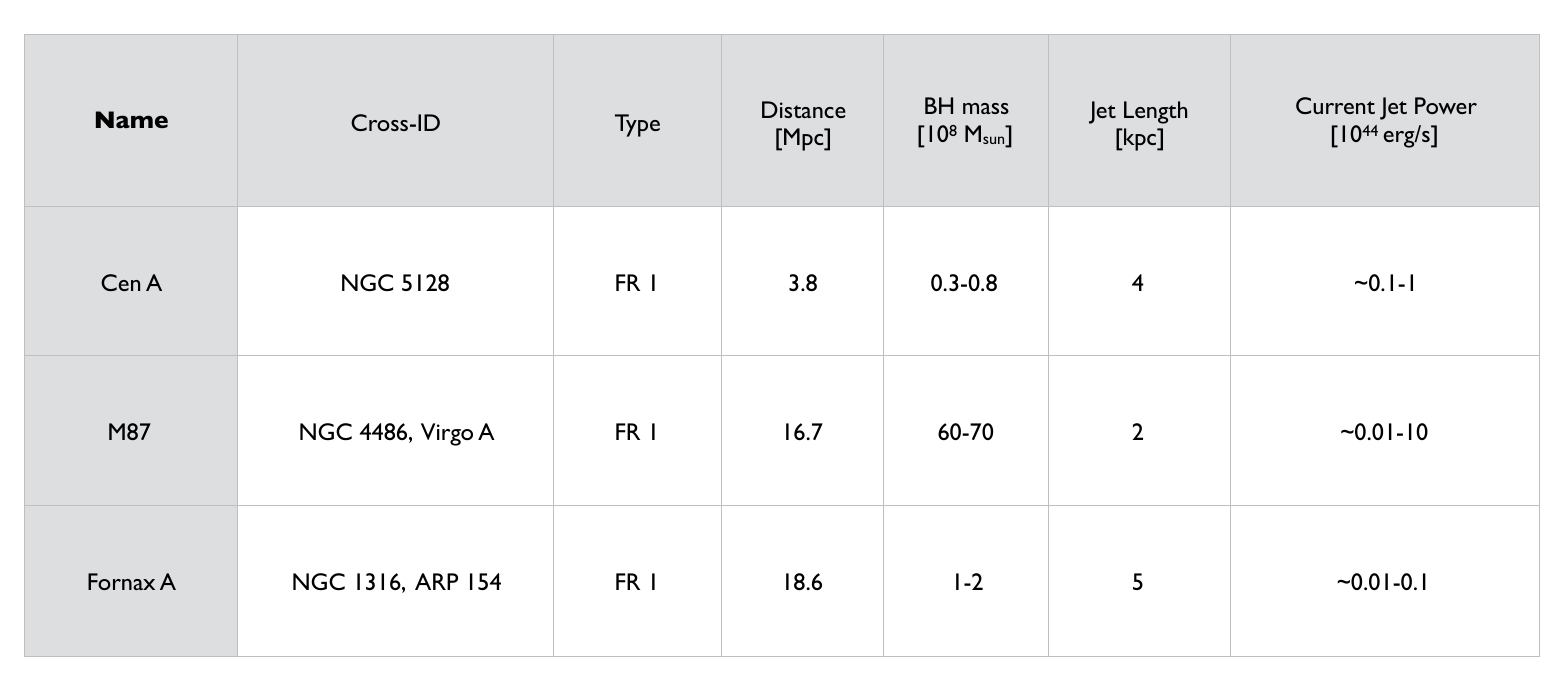}
\caption{The three nearest radio galaxies, considered to be UHECR prime candidate sources, along 
with values for their black hole mass, typical (projected) jet lengths and estimates for the "current" jet 
power. The latter are somewhat uncertain and model-dependent. Cross-IDs give their alternative 
source identifications}\label{rg_list}
\end{center}
\end{figure}

Radio-loud active galaxies have for long been discussed as potential sources of the highest energy 
cosmic rays, e.g., refs~\cite{Brecher1972,Hillas1984,Takahara1990,Rachen1993,Blandford2000,
Rieger2009a,Hardcastle2010,Biermann2012,Matthews2018}. Although active galaxies form only a 
small fraction ($\sim 1-3\%$) of all galaxies, they are known as powerful, non-thermal emitting sources. 
Radio-loud Active Galaxies (or jetted AGNs, cf.~\cite{Padovani2017}) are in fact characterised by 
multiple sites of efficient electron and putative cosmic-ray acceleration, from the vicinity of the black 
holes ($r_g$) up to multi kiloparsec-scale relativistic jets and giant lobes. 
While originally the more powerful Fanaroff-Riley (FR) II radio galaxies have been considered as promising 
UHECR accelerators, more recently the FR~I radio galaxies Cen A, M87 and Fornax A have also been 
in the focus of attention as they are nearby and come with radio fluxes about an order of magnitude higher 
than other nearby radio galaxies, see also Figs.~\ref{local_RG}-\ref{rg_list}.\footnote{Characteristic (local) number 
densities are of the order $n\sim 10^{-5}-10^{-4}$ Mpc$^{-3}$ for FR~I and $n\sim 10^{-8}-10^{-7}$ 
Mpc$^{-3}$ for FR~II radio galaxies, cf. \cite{Padovani2011,Prescott2016}. Cen~A is the closest ($d\sim 
3.8$ Mpc) FR~I, while Pictor~A is one of the closest ($z=0.035$, $d\sim 150$ Mpc) FR~II radio galaxies. 
There are not many FR~II sources known within the GZK horizon.} For example, within $\sim 130$ Mpc, 
a fraction of about $\sim 0.6\%$ galaxies in the 2MRS catalog have been identified as radio-loud; out of 
those only 24 galaxies have an intrinsic radio (1 GHz) luminosity greater or equal to Cen~A, most of them 
being much more distant ($>10$ times) than Cen~A \cite{VanVelzen2012}.  

For comparison and discussion, the relevant physics characteristics of the three nearest radio sources 
are, in the following, shortly summarized (cf. also Figure~\ref{rg_list}):
\begin{itemize}
\item{\bf Cen A:} As the closest active galaxy to Earth ($d\simeq 3.8$ Mpc) Centaurus A (Cen~A) is one of 
the most prominent radio sources in the southern sky. Its inner region harbours a warped, relatively massive 
dust and gas disk, most likely the product of a past merger activity between a small, gas-rich spiral galaxy 
and a larger, elliptical one. There are in fact several lines of evidence suggesting that at least one major 
merger event occurred some $10^8-10^9$ years ago \cite{Israel1998}. Over time, Cen~A then most likely 
underwent several phases of AGN activity \cite{Saikia2009,Struve2010}. Its central engine hosts a black 
hole of mass $M_{\rm BH} \simeq (0.3-0.8) \times 10^8 M_{\odot}$ \citep[][]{Neumayer2010}, and 
currently emits a (nuclear) bolometric (photon) luminosity of $L_{\rm bol} \sim 10^{43}$ erg/sec \cite{Whysong2004}. 
At radio frequencies, Cen~A is known for its peculiar morphology including a compact radio core, a sub-pc 
scale jet and counter-jet, a one-sided kpc-scale jet ($\sim 4$ kpc in projection), inner radio lobes as well as
giant outer lobes with a length of several hundreds of kiloparsec, e.g. \cite{Feain2011}. At high photon energies, 
gamma-ray observations have revealed GeV emission from the giant lobes as well as TeV emission along 
its large-scale jet, demonstrating the operation of efficient in-situ particle acceleration \cite{Fermi_CenA2010,
Yang2012,HESS_CenA2020}. VLBI observations indicate that Cen~A is a non-blazar source, with its inner 
jet misaligned by about $(12-45)^{\circ}$ and characterized by moderate bulk flow speeds $\leq 0.5$ c for 
the radio region probed \cite{Mueller2014}. EHT observations have revealed asymmetric (spine-sheath type) 
jet emission on scales of a few hundred Schwarzschild radii \cite{Janssen2021}. On larger scales ($>100$ pc), 
trans-relativistic jet speeds of $\sim 0.5-0.7$ c have been inferred \cite{Hardcastle2003,Snios2019}. 
Estimates for the "current" kinetic jet power are in the range $L_j \sim 10^{43}-10^{44}$ erg/s \cite{Yang2012,
Wykes2015,Sun2016}.\\  

\item{\bf M87:} As the second closest active galaxy ($d\simeq 16.4$ Mpc \citep{Bird2010}), the Virgo Cluster 
galaxy M87 (NGC 4486) hosts one of the most massive black holes of $M_{\rm BH} \simeq (6.5\pm0.7) \times 
10^9\,M_{\odot}$ \cite{EHT2019,Broderick2022}. Classified as low-excitation FR I source, M87 is thought 
to be currently accreting in a radiatively inefficient (RIAF) mode \cite{Reynolds1996,EHT2019b}. Common 
estimates for its total nuclear (disk and jet) bolometric luminosity in fact do not exceed $L_{\rm bol} \simeq 
10^{42}$ erg/s by much \cite{Biretta1991,Owen2000,Whysong2004,Prieto2016}. In the radio, M87 exhibits an 
extended morphology including a sub-pc scale jet and counter-jet, a one-sided kpc-scale jet ($\sim 2$ kpc in 
projection) that is also seen in X-rays, inner lobes as well as a large-scale halo of size $\sim 50$ kpc 
\cite{Biretta1991,Owen2000,Marshall2002,deGasperin2012,Walker2018}. The X-ray emission seen in the 
large-scale jet of M87 provides evidence for the presence of ultra-relativistic electrons and the operation 
of efficient in-situ acceleration \cite{Sun2018}. Both, on sub-parsec as well as on kpc scales evidence has 
been found for a stratified jet structure compatible with a fast spine and a slower sheath \cite{Perlman1999,
Mertens2016}. There is evidence, including superluminal motion seen at optical and X-ray energies, that 
on larger scales ($>100$ pc) the jet still possesses significantly trans-relativistic ($\gppr 0.7$ c) speeds 
\cite{Meyer2013,Snios2019b}.
Estimates for the kinetic power of the jet in M87 are somewhat uncertain and range from $10^{42}$ to
$10^{45}$ erg/s \cite{Reynolds1996,deGasperin2012,Prieto2016,Sun2018}. 
Being at the center of a galaxy cluster, it has been argued that M87 might be repeatedly fed by accretion 
of satellite galaxies containing intermediate mass black holes \cite{Safar2019}. There is evidence that 
accretion of a smaller galaxy has caused an important modification of its outer halo in the last Gyr 
\cite{Longo2015}. On the other hand, many X-ray features (e.g., cavities) on halo-scale appear to be 
a result of recurrent AGN activity (on timescales probably down to $\sim 10^7$ yr) \cite{Forman2005,
Forman2017}.\\

\item{\bf Fornax A:} Located at $d\simeq 20.8$ Mpc \cite{Cantiello2013}, Fornax A (NGC 1316) is 
the brightest member of the Fornax Cluster and a prominent merger remnant. In the optical, it resembles 
an elongated spheroid-type galaxy. In the radio, an S-shaped nuclear radio jet ($\sim 5$ kpc in projection) 
and giant double lobes spanning $\sim 350$ kpc are seen \cite{Geldzahler1984}. Fornax A harbours a 
supermassive black hole of $\simeq(1.5 \pm0.8)\times 10^8 M_{\odot}$ \cite{Nowak2008}. 
Given its peculiar morphology, Fornax A is thought to be shaped by a major merger about 3 Gyr ago, 
probably followed by some minor mergers of smaller companions \cite{Mackie1998,Goudfrooij2001,
Iodice2017}. There is evidence that the extended radio lobes have been formed by multiple episodes 
of nuclear activity on timescales of several Myr \cite{Maccagni2020}. Characteristic estimates for the 
current jet power are in the range $L_j \sim 10^{42-43}$ erg/s \cite{Balmaverde2008,Russell2013,
Maccagni2020}. While Fornax A is commonly classified as an FR~I source, different episodes of 
activity could explain some challenges as to a unique classification, e.g., the phase that shaped 
the lobes appears to have been more powerful compared with the activity of the central 
emission~\cite{Maccagni2020}.\\
\end{itemize}

As shown below, jet power estimates can provide important constraints for UHECR acceleration 
(Sec.~5). Results for the current jet power are, however, not necessarily conclusive in this regard. 
In particular, as UHECRs are confined to magnetic fields, they might escape only slowly from the 
AGN environment. Since most likely many radio sources have undergone several phases of AGN
activity, current FR~I sources could well have seen more powerful jets in the past. If this is the 
case, then the history of a source becomes relevant for understanding its role as a possible 
UHECR accelerator. It has been argued, for example, that this particularly applies to Cen~A and 
Fornax A where UHECR particles, accelerated in the past, could still be escaping from their giant 
lobes \cite{Matthews2018,Matthews2019b}. Note that the presence of strong magnetic fields on 
cluster scales might also suppress the escape of nuclear UHECRs from the Virgo Cluster (M87) 
\cite{Biteau2022b}.\\ 
For an informative, recent discussion of the role of individual FR~I radio galaxies as possible 
UHECR sources the reader is also referred to refs.~\cite{Eichmann2018,Wykes2018} (Cen A), 
\cite{Kobzar2019} (Virgo A/M87), \cite{Fraija2019} (Cen B) and \cite{Matthews2018} (Fornax A), 
respectively.

\section{Basics Physics Constraints on UHECR candidate sources}\label{constraints}
In general, to be considered as a realistic astrophysical candidate, a possible UHECR source has 
to satisfy a few basic physics requirements. For example, in order to be capable of accelerating 
charged particles to ultra-high energies $E$, the relevant particles need to remain confined within 
the accelerator, i.e., the particle Larmor radius $r_L \simeq E/(ZeB)$ has to remain smaller than the 
characteristic dimension $L$ of the source ($r_L \leq L$) , where $B$ is the source magnetic field 
strength and $Z$ the charge number. This introduces a basic (Hillas) bound
\begin{equation}\label{Hillas}
E \leq 10^{20} \,Z\, \left(\frac{B}{10\mu\,\mathrm{G}}\right) \left(\frac{L}{10~\mathrm{kpc}}\right)\;\mathrm{eV}\,,
\end{equation}  
\begin{figure}[h]
\vspace*{0cm}
\begin{center}
\includegraphics[width=0.8 \textwidth]{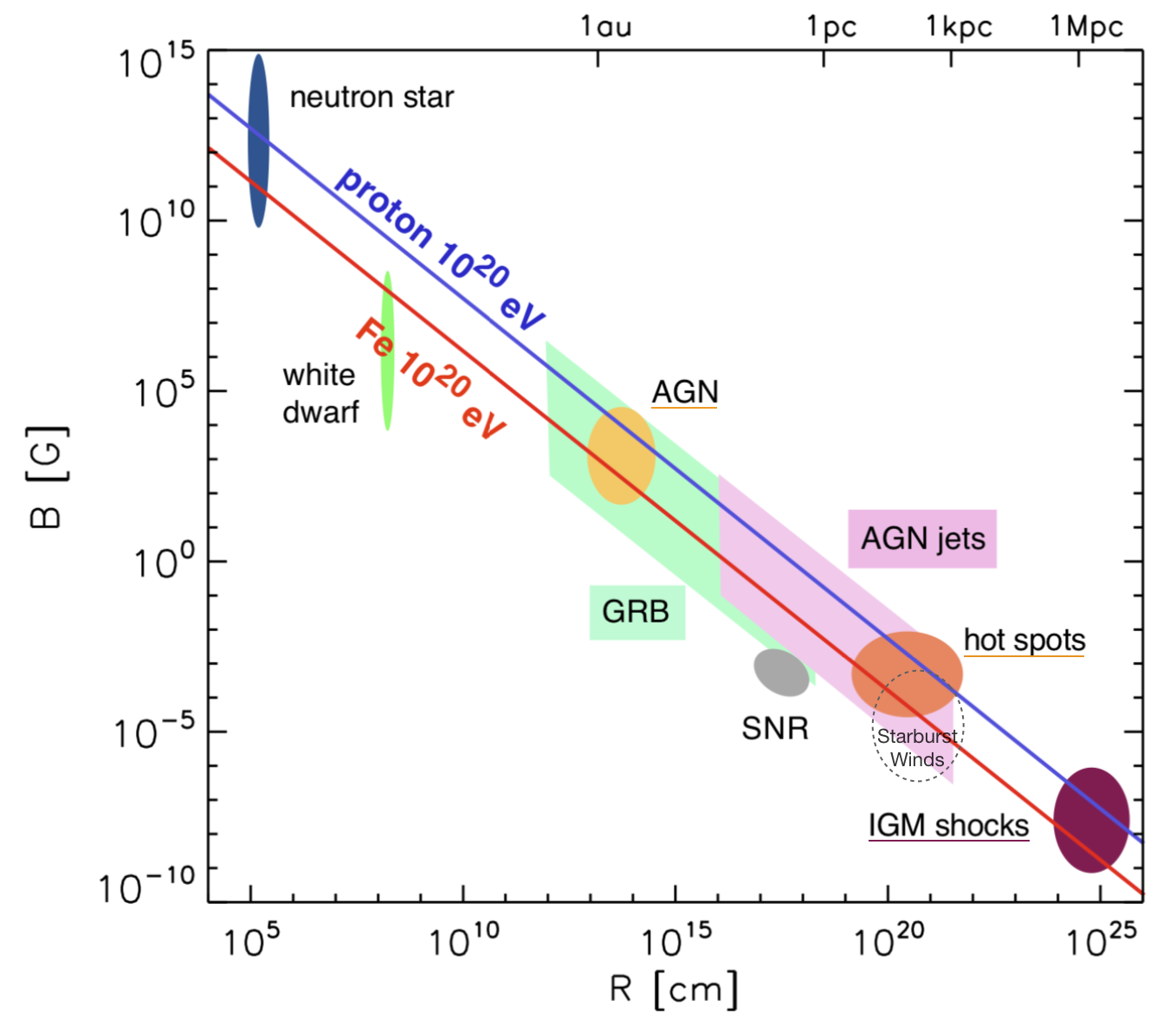}
\caption{Updated Hillas diagram \cite{Hillas1984}, obtained by requiring the Larmor radius for a $10^{20}$ 
eV particle to be smaller than the characteristic source size $R$. The blue line and the red line show the
limits for protons and iron nuclei, respectively. Also shown are astrophysical objects that can meet the 
requirements on source size and magnetic field $B$ (above the respective lines). AGN (radio galaxies) 
typically represent suitable environments. Adapted based on ref.~\cite{Kotera2011}.}
\label{Hillas_fig}
\end{center}
\end{figure}
which is often visualised by means of a $B$-$L$ (or $R$) diagram, Figure~\ref{Hillas_fig}, delineating 
potential astrophysical sources that meet this requirement. Radio galaxies typically represent suitable 
candidates in this regard. An improved Hillas constraint is obtained once characteristic astrophysical 
flow velocities, $V$, are properly taken into account. This relates to the fact that particle acceleration 
will be constrained by the accessible, motional electric field, i.e., $\vec{\varepsilon}= - (\vec{V}/c) \times 
\vec{B}$, which gives
\begin{equation}\label{Hillas_2}
E \leq Z e\, \varepsilon \,L = 10^{20}\, Z\, \beta \,\left(\frac{B}{10\mu\mathrm{G}}\right) 
\left(\frac{L}{10~\mathrm{kpc}}\right)\;
\mathrm{eV}\,,
\end{equation} where $\beta =V/c$. Accordingly, efficient UHECR acceleration generally requires source
environments with fast flow speeds $\beta$, strong magnetic fields and/or large volumes \cite{Hillas1984}. 
Obviously, the requirements are significantly relaxed if the composition at the highest energies would be 
sufficiently heavy (e.g., $Z\geq 10$).\\
In general, the basic constraints above only represent necessary but not sufficient conditions for UHECR 
acceleration. Applications and more refined constraints have been discussed by various authors, e.g.,
refs.~\cite{Norman1995,Blandford2000,Aharonian2002,Lemoine2009}. In particular, particle acceleration 
needs to proceed on a characteristic timescale ($t_{\rm acc}$), which remains faster than both, the 
escape ($\tau_{\rm esc}$) and the relevant radiative loss ($\tau_{\rm loss}$) timescale, respectively. 
Since $t_{\rm acc}$ depends on acceleration physics, and $\tau_{\rm esc}$, $\tau_{\rm loss}$ on individual 
source physics, one important outcome of this is that detailed "multi-messenger" studies of individual sources 
(allowing to, e.g., infer magnetic fields and flow speeds) become highly relevant again. Hence, while an 
individual source may formally belong to the "right" class of objects in the Hillas diagram, its true suitability 
can only be verified within a wider astrophysical perspective. To illustrate this, note that already the three 
closest FR~I radio galaxies reveal a significant spread in parameters, cf. Figure~\ref{rg_list}.\\
The Hillas source constraint can be used to infer the magnetic power $L_B$ a putative accelerator needs 
to carry to allow for UHECR production. For collimated, non-relativistic ($V \ll c$) jetted outflows for example, 
$L_B= 2\pi R^2 u_B V$. With $u_B=B^2/8\pi$, $R\simeq L$ and $L$ from eq.~(\ref{Hillas_2}), one obtains
\begin{equation}\label{Hillas_power}
L_B \simeq 8\times 10^{44} \left(\frac{1}{\beta}\right) \left(\frac{E/Z}{10^{20}~\mathrm{eV}}\right)^2 \,
                    \mathrm{erg/sec}\,.
\end{equation} 
This can be extended to shock-type particle acceleration in relativistic outflows, see e.g. 
ref.~\cite{Lemoine2009}, resulting in a strong, generalized Hillas constraint on the required source 
power. The argument utilises (i) the requirement that particle acceleration must proceed faster than 
dynamical escape, i.e., that in the local, co-moving frame $t_{\rm acc}' < t_{\rm dyn}'$ (primed quantities 
referring to the co-moving frame). 
If one expresses the timescale for shock acceleration in terms of the gyro-timescale, i.e., $t_{\rm acc}' 
=\eta\,r_L'/c  = \eta\,E'/(Z eB'c)$, where $\eta\geq 1$ provides a measure for the efficiency, and uses 
that the dynamical timescale can be written as $t_{\rm dyn}' = d/(\Gamma\,V)$ (with $d$ the longitudinal 
length scale and $\Gamma$ the flow Lorentz factor), then maximum achievable UHECR energies in 
the observer's frame can be re-expressed as $E = \Gamma E' \leq Z e B' d /(\beta \eta)$. 
To allow for this, however, the outflow (ii) would need to carry a high enough magnetic power $L_B 
= 2\pi r^2 \Gamma^2 u_B' V$, where $r$ denotes the lateral half width and $u_B' \equiv B'^2/8\pi$. Using 
the $E-B'$ relation from (i), and noting that we can approximate $r \sim \theta_j d$, where $\theta_j$ is 
the jet/flow half opening angle, this can be rewritten as $L_B \geq \theta_j^2 \Gamma^2 \eta^2 E^2 
\beta^3 c/(2\, Z\, e)^2$, or
\begin{equation}\label{lower_limit}
L_B \gppr 8 \times 10^{44} \theta_j^2 \Gamma^2 \eta^2 \beta^3 \left(\frac{E/Z}{10^{20}~\mathrm{eV}}\right)^2
                    \, \mathrm{erg/sec}\,. 
\end{equation} For the commonly assumed $\theta_j \sim 1/\Gamma$, this thus implies a lower limit on the 
required source luminosity for a steady relativistic ($\beta \sim 1$) outflow to accelerate charged particles to 
energies $E$, of $L_B \gppr 8\times 10^{44}\, Z^{-2}\, (E/10^{20}\mathrm{eV})^2$ erg/sec. For non-relativistic 
shocks, where $\eta \sim \beta^{-2} = (c/V_s)^2$, this condition becomes even more stringent as $L_B \propto 
1/\beta$, cf. eq.~(\ref{Hillas_power}).
In general, the situation is again much relaxed if a heavier ($Z\geq 10$) composition would prevail at the 
highest end, as suggested by the experimental findings reported in Sec.~\ref{sec_composition}. In particular, 
the three closest FR~I radio galaxies, Cen~A, M87 and Fornax A could then, already in their current phase 
of activity, satisfy the power requirement imposed by shock acceleration, cf. Figure~\ref{rg_list}. On the other 
hand, since UHECRs get deflected in (Galactic) magnetic fields by $\theta \sim d/r_g \propto (Z/E)$, one 
may have to address, why, assuming (!) a conventional Fermi-type particle spectrum, we do not seem to 
observe a (more) pronounced anisotropy associated with protons at $E/Z$, cf. 
refs.~\cite{Lemoine2009,Liu2013,Lemoine2018}.\\
Note that from an astrophysical point of view, the power requirement, eq.~(\ref{lower_limit}), disfavours 
starburst galaxies (SBGs) as promising UHECR accelerators. Since the winds in SBGs are comparatively 
slow (shock velocities less than a few $\sim 1000$ km/s) and carry rather modest power $L_{\rm SBG} 
\sim 10^{42}$ erg/sec, cf. refs.~\cite{Heckman2017,Romero2018,Zhang2018,Matthews2018}, shock-driven
UHECR acceleration in SBGs is unlikely to be efficient. Unless SBGs would be (more) frequent hosts of 
explosions of UHECR progenitors that match the Hillas criterion (e.g., long GRBs), this would suggest 
that a putative UHECR-SBG correlation (if any) is more consistently interpreted as being caused by a 
scattering of UHECRs originally accelerated in radio galaxies (see 3.4.).

\section{UHECR Acceleration in AGNs: Sites and Mechanisms}
\begin{figure}[h]
\vspace*{0cm}
\begin{center}
\includegraphics[width=0.45 \textwidth]{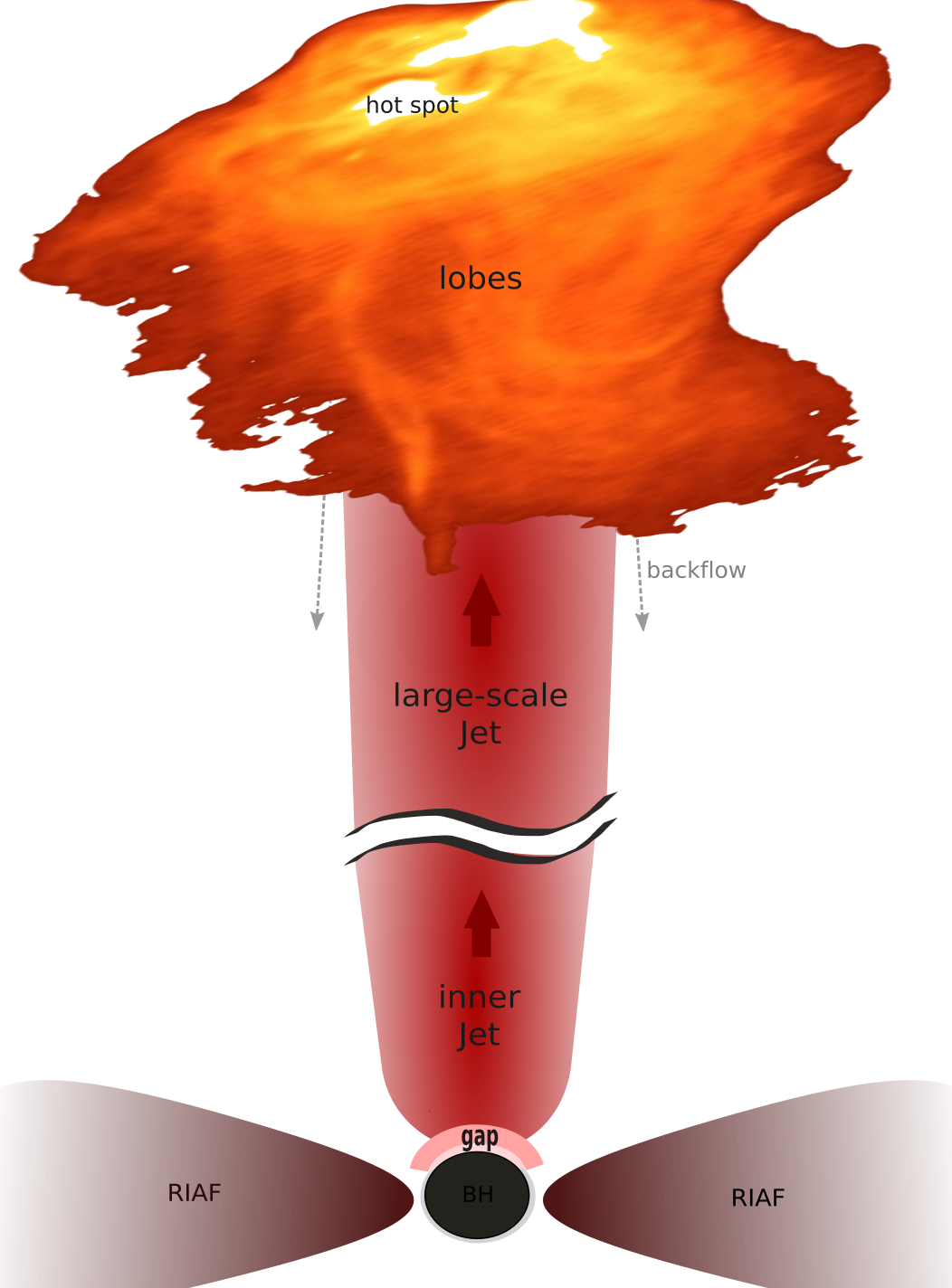}
\caption{Sketch: Radio galaxies or jetted AGNs provide multiple sites for cosmic particle acceleration, 
from the black hole vicinity (gaps) to their large-scale lobes. The preferred acceleration mechanism 
can change, dependent on the location within the source. Some of the mechanisms might as well 
work in tandem as to UHECR production.}
\label{AGN_model}
\end{center}
\end{figure}

\begin{figure}[h]
\vspace*{0cm}
\begin{center}
\includegraphics[width=0.8 \textwidth]{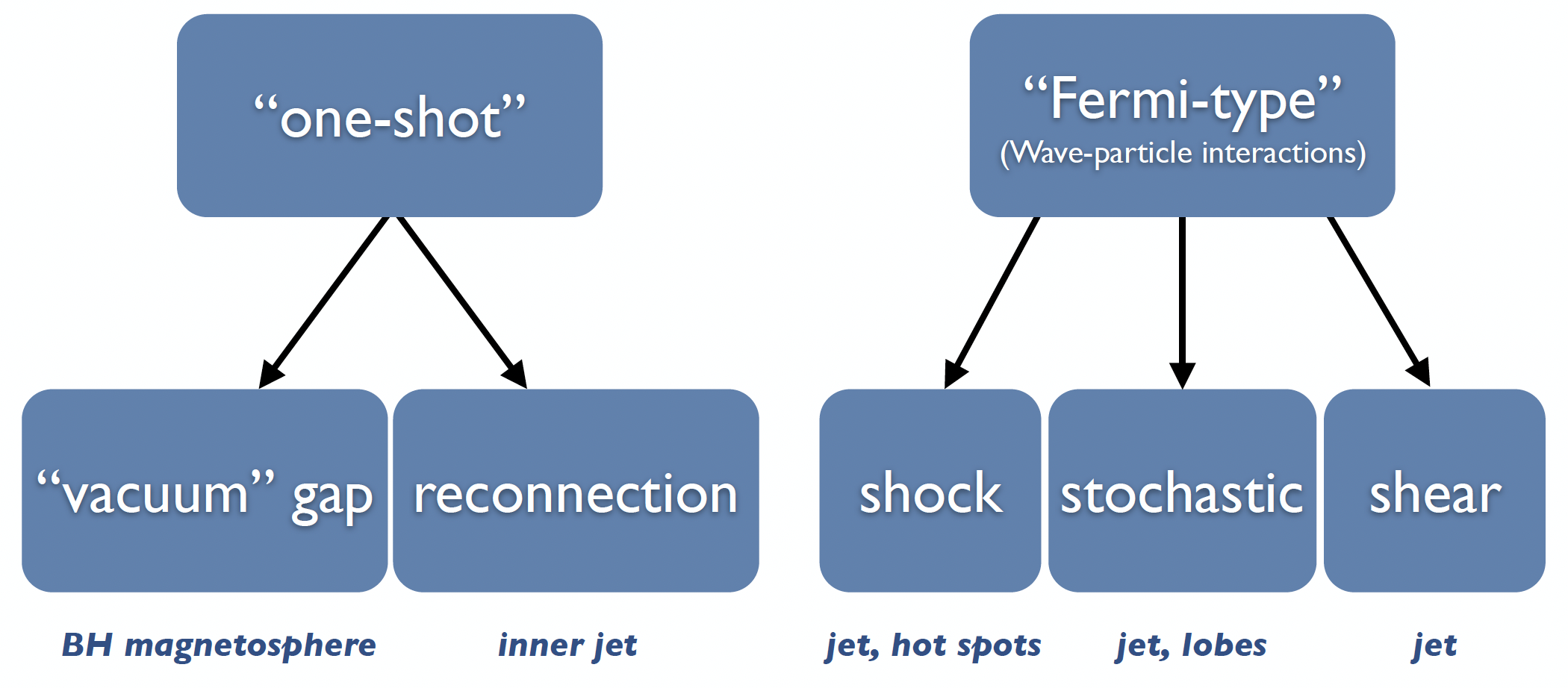}
\caption{Possible acceleration mechanisms and preferred sites in AGNs (not exhaustive).
Sometimes the processes could be intertwined, e.g., reconnection could also be a driver 
for part of the turbulence required for stochastic acceleration.}
\label{acceleration}
\end{center}
\end{figure}

Radio-loud or jetted AGNs provide multiple sites for efficient particle acceleration, including the vicinity
of the black hole (in so-called "vacuum" gaps), the inner (pc-scale) and outer (kpc-scale) jet, the jet 
termination shock ("hot spot") in the case of powerful sources, the back-flowing region and the 
jet-inflated, large-scale lobes, see e.g. Figure~\ref{AGN_model}.
In principle, a variety of acceleration mechanisms can be operative, from e.g. "one-shot" 
type particle acceleration (by an emergent, ordered electric field component) to Fermi-type 
particle acceleration \cite{Rieger2007,Lemoine2019} based on particles repeatedly scattering 
off moving magnetic inhomogeneities (plasma waves) and drawing energy from the velocity 
changes they experience during their diffusion process, cf. Figure~\ref{acceleration}.
The operating mechanism can change, dependent on the location within the source.
In the following, three exemplary, recent scenarios will be highlighted. 

\subsection{Cosmic-Ray Acceleration in the Magnetospheres of Rotating Supermassive Black Holes}
\begin{figure}[htbp]
\begin{center}
\includegraphics[width = 0.5 \textwidth]{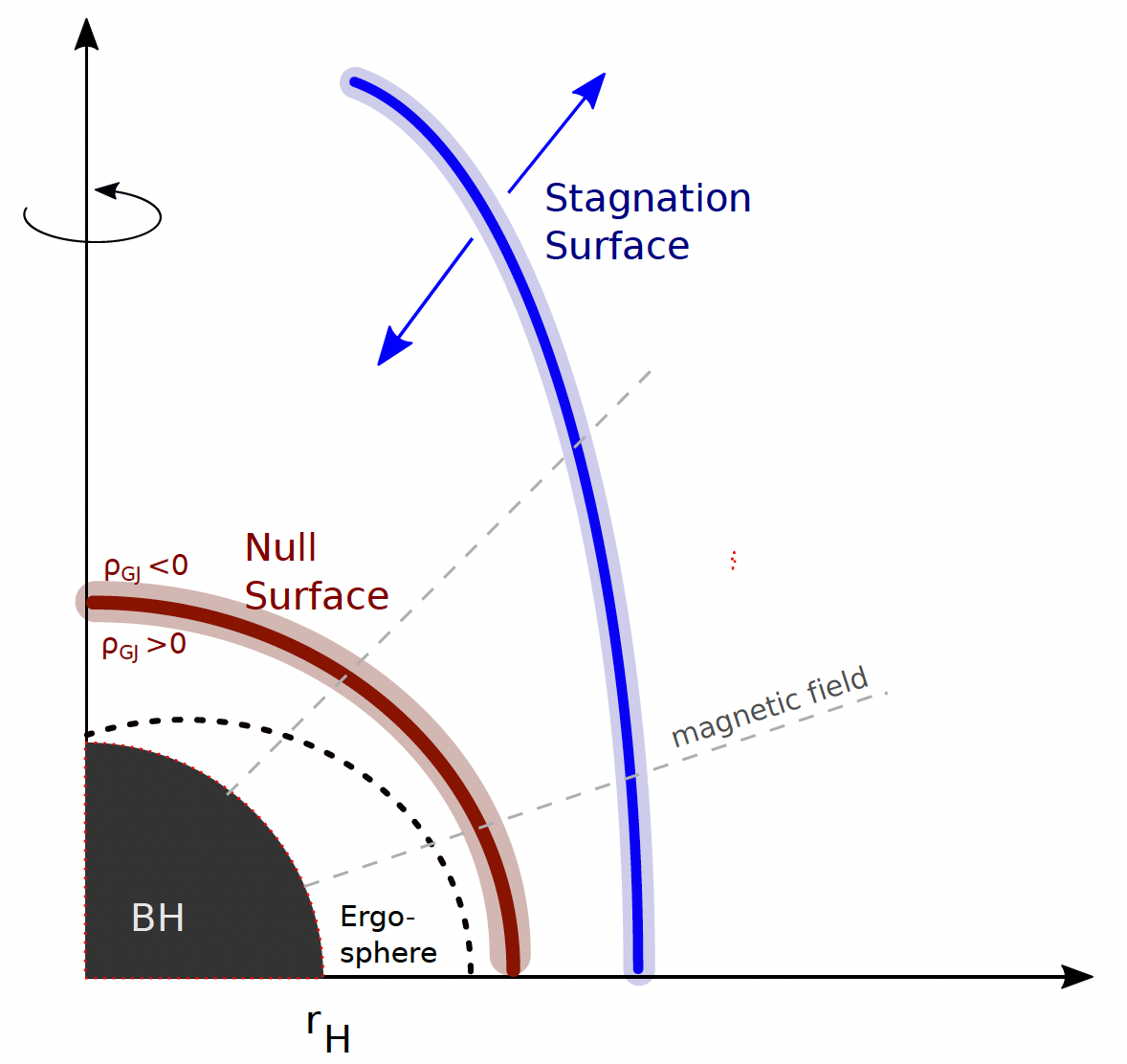}
\caption{Illustration of possible locations of charge-deficient regions (gaps) around a rotating black 
hole in which UHECR acceleration could occur. The red line denotes the null surface across which 
the required Goldreich-Julian charge density $\rho_{\rm GJ}$ changes sign, while the blue line 
delineates the stagnation surface from which stationary MHD flows start.}
\label{gap}
\end{center}
\end{figure}

\begin{figure}[htbp]
\begin{center}
\includegraphics[width = 0.90 \textwidth]{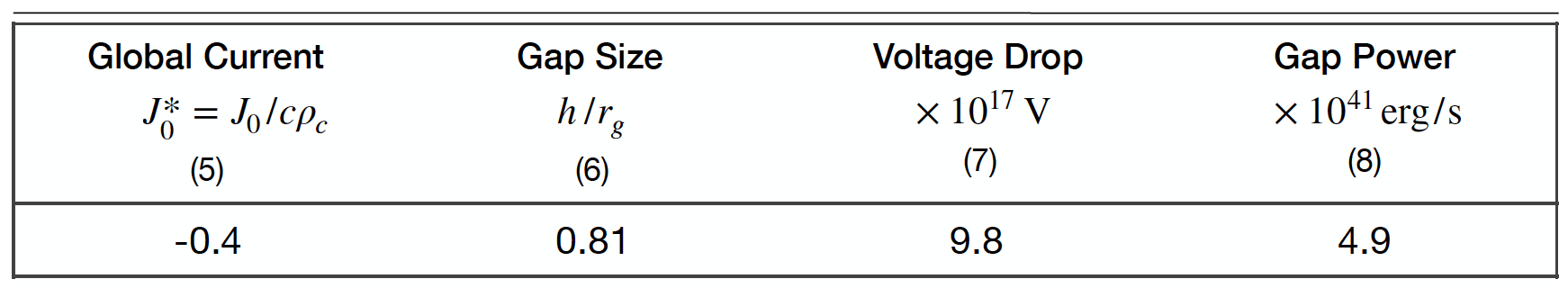}
\caption{Characteristic gap properties inferred for M87 assuming $M_{\rm BH}=6.5\times 10^9 
M_{\odot}$ and $\dot{m} = 10^{-5.75}$. The columns give the extension or width $h$ of the gap, 
the associated voltage drop, and the total gap power for a global current $J_0^* = -0.4$. 
Based on ref.~\cite{Katsoulakos2020}.}
\label{gap_properties_M87}
\end{center}
\end{figure}

The vicinity of accreting, supermassive black holes is permeated by strong magnetic fields $B$ whose 
strengths can reach thousands of Gauss. Fields that are dragged into rotation ($\Omega_F$) 
by the black hole can induce an electric field, that, if not screened, can facilitate efficient particle 
acceleration. The possibility of UHECRs acceleration in the magnetospheres of supermassive black 
holes has been discussed by several authors, see e.g. refs.~\cite{Boldt1999,Levinson2000,
Neronov2009,Rieger2011,Ptitsyna2016,Moncada2017,Katsoulakos2020}. In this framework, particle 
acceleration essentially relies on the emergence of an electric field component parallel to the magnetic 
field in a charge-deficient ("gap") region close to a rotating black hole, cf. also ref.~\cite{Rieger2018} 
for a related review. This could occur either around the so-called null surface across which the 
Goldreich-Julian charge density, $\rho_{\rm GJ}\sim (\Omega_F- \omega) B$, required to screen 
the electric field, changes sign, or at the stagnation surface that separates MHD in- and outflows, 
see Figure~\ref{gap}.
The maximum available voltage drop for a gap of width $h$ around a rotating (Kerr) black hole is 
of the order of \cite{Katsoulakos2018} 
\begin{equation}\label{voltage_drop}
\Delta \Phi = \frac{1}{c} \Omega_F r_H^2 B_H \left(\frac{h}{r_g}\right)^2 
                     \simeq 2\times 10^{21} \dot{m}^{1/2} M_9^{1/2} \left(\frac{h}{r_g}\right)^2\,\mathrm[V]\,,
\end{equation} assuming a horizon-threading magnetic field $B_H \simeq 2\times 10^5 \dot{m}^{1/2} 
M_9^{-1/2}$ G, a black hole of mass $M_9 \equiv M_{\rm BH}/10^9 M_{\odot}$, field line rotation 
with $\Omega_F= \Omega_H/2=c/4r_g$, normalized accretion rate $\dot{m} \equiv \dot{M}/
\dot{M}_{\rm Edd}$ and a gravitational radius $r_g \equiv GM_{\rm BH}/c^2$. This would seem to 
suggest that in massive ($M_{\rm BH} \gppr 10^9 M_{\odot}$) and active ($\dot{m}\gppr 10^{-3}$) 
galaxies, ultra-high energies $E= Z\,e\,\Delta\Phi \sim 10^{20}Z$ eV might be accessible, provided
no other limitations exist. However, charged particles moving along curved magnetic field lines 
will also undergo curvature radiation losses. Taking these radiative losses into account introduces a 
general upper limit for a particle of mass $m$ and charge number $Z$, which is of the order of
\cite{Katsoulakos2018}
\begin{equation}\label{curvature}
 E_{\rm max} \lppr 10^{19} \,\frac{\dot{m}^{1/8}}{Z^{1/4}} \,M_9^{3/8} \left(\frac{h}{r_g}\right)^{1/4}
                                   \left(\frac{m}{m_p}\right)\;\mathrm{eV}
\end{equation} for curvature radii of the order of the gravitational one. Note that for low accretion rates 
and narrow gaps, eq.~(\ref{voltage_drop}) will be more constraining and set the relevant limit. 
Note also, that eq.~(\ref{curvature}) provides an example for a process where the maximum energy 
$E_{\rm max}$ no longer scales linearly with $Z$.\\ 
In reality, the prospects for UHECR acceleration in black hole magnetospheres are expected to be
further reduced: Unless accretion rates are sufficiently low (i.e., well below $\dot{m} \sim 10^{-4}$), 
the ambient soft photon environment will facilitate efficient pair production in the gap (if gaps form 
at all, cf.~\cite{Levinson2011,Hirotani2016}), leading to gap sizes $h \ll r_g$, thereby significantly 
reducing the particle energies achievable, cf. eq.~[\ref{voltage_drop}]. On the other hand, if accretion 
rates would be very low, $\dot{m}\ll1$, as in the case of "quiescent quasars", externally supported 
magnetic fields ($B_H \propto \dot{m}^{1/2}$) would be much reduced as well. 
By hosting one of the most massive black hole ($M_9=6.5$) and accreting at low rates (RIAF mode), 
the radio galaxy M87 represents an instructive example to explore this in more details this. 
\begin{figure}[h]
\begin{center}
\includegraphics[width = 0.60 \textwidth]{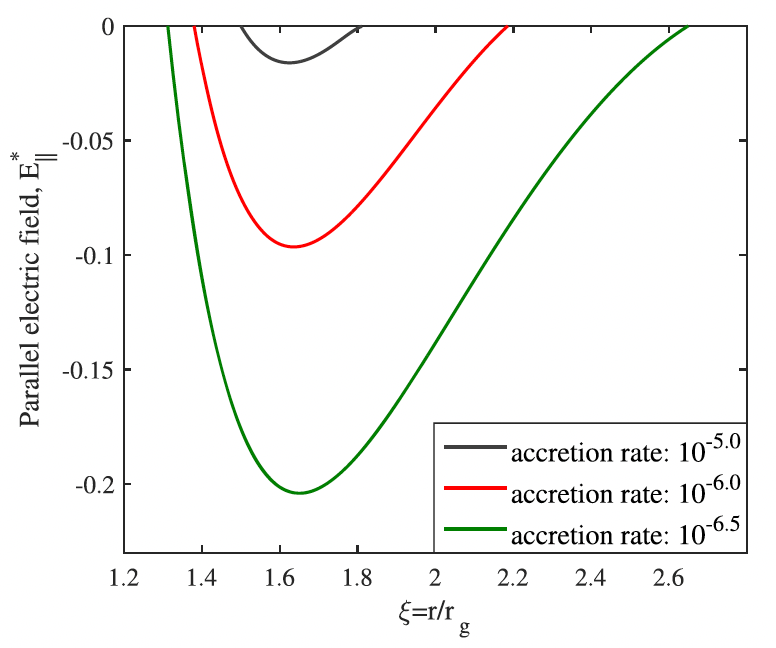}
\caption{Characteristic gap evolution for a rotating supermassive black hole of mass $10^9 M_{\odot}$ 
that accretes in a radiatively inefficient mode. The curves show the radial distribution of the parallel 
electric field component for different accretion rate. The gap width and voltage drop increase as the 
accretion rate drops because pair creation of energetic photons with the soft photon field becomes 
less efficient. Results like these can be used to evaluate the gap properties and prospects for UHECR 
acceleration for a particular source of interest. From ref.~\cite{Katsoulakos2020}.}
\label{gap_electric_field}
\end{center}
\end{figure}
Solving the system of relevant partial differential equations (e.g., Gauss' law, equation of particle 
motion, continuity equation), one can seek for self-consistent steady gap solutions \cite{Hirotani2016,
Katsoulakos2020}, that provide insights into characteristic electric field strengths and achievable 
voltage drops, see e.g. Figure~\ref{gap_electric_field}.
For example, taking an accretion rate $\dot{m}=10^{-5.75}$, which is close to the mean MAD value 
used in Event Horizon Telescope (EHT) GRMHD simulations for M87 \cite{EHT2019b} and which 
corresponds to jet powers of a few times $10^{43}$ erg/s, the gap properties for M87 can be 
determined, see Figure~\ref{gap_properties_M87}. These gap solutions imply a voltage drop for M87 
which is of the order of $\sim 10^{18}$ V, suggesting that its black hole is not an efficient UHECR 
proton accelerator.
These considerations reinforce the belief that in our sources of interest, efficient gap-type particle 
acceleration of protons to energies much beyond $E \sim 10^{18}$ eV is problematic. We note that 
since efficient UHECR production in the black hole magnetosphere is accompanied by VHE 
curvature emission, gamma-ray observations could in principle allow to further constrain UHECR 
acceleration in under-luminous or "passive" AGNs, e.g. \cite{Levinson2000,Pedaletti2011}.

\subsection{Shear Particle Acceleration in the Large-Scale Jets of AGNs}
The large-scale jets in AGNs are efficient electron accelerators as evidenced by the detection of 
extended, non-thermal X-ray emission along their kpc-scale jets. The currently most favoured 
mechanism for producing the observed X-ray emission is electron synchrotron radiation, see e.g. 
ref.~\cite{George2016,Breiding2017} for a discussion. In M87, the radio-X-ray spectral characteristics 
in fact preclude an inverse Compton (CMB up-scattering by low energy electrons) origin \cite{Sun2018}, 
while in the case of Cen~A the recent detection of extended VHE emission (due to up-scattering of 
dust by very energetic electrons) has verified the X-ray synchrotron interpretation and demonstrated
the presence of ultra-relativistic electrons with Lorentz factors $\gamma_e \gppr 10^8$ along its
kpc-scale jet \cite{HESS_CenA2020}. Given the short synchrotron cooling timescale $t_{\rm cool}
\propto 1/\gamma_e$ and cooling length, $c t_{\rm acc} \ll 1$ kpc, of these electrons, the operation 
of a distributed acceleration mechanism is required to keep electrons energized and to account for 
the extended emission. Stochastic (2nd order Fermi) and shear particle acceleration are among the 
most promising acceleration mechanisms in this regard \cite{Liu2017,Tavecchio2021,Rieger2021,
Wang2021}, cf. Figure~\ref{acceleration}. It seems then natural to suppose that these mechanisms 
could also facilitate the acceleration of protons and hadrons to high energies. Given the reduced 
energy losses and photo-disintegration, large-scale jets may in fact represent the most promising 
sites of UHECR acceleration. As has been argued in Sec.~\ref{constraints}, however, to adequately 
assess the UHECR potential of a given source eventually requires a wider, multi-messenger approach. 
Insights from multi-wavelength SED modelling, for example, can be used to constrain key physical 
parameters of the jet (e.g., magnetic field strength, flow speed), e.g. \cite{Wang2021}, and thereby 
inform our understanding. 

In general, AGN jets are expected to exhibit some internal jet stratification and velocity shearing. 
A fast-spine, slower-sheath structure, for example, may already be introduced close to the black 
hole, and be related to an ergospheric-driven (Blandford-Znajek) jet surrounded by a disk-driven 
(Blandford-Payne) outflow, cf. ref.~\cite{Mizuno2022} for a review. 
As the jet continues to propagate, instabilities and mixing will continue to modify its shape and 
bulk velocity structure \cite{Perucho2019a}. Limb-brightened features, different velocity and 
polarization pattern, for example, are observable signatures typically attributed to such flow 
structures, see e.g., refs.~\cite{Mertens2016,Kim2018}. 

\begin{figure}[htb]
\begin{center}
\includegraphics[width=0.45 \textwidth]{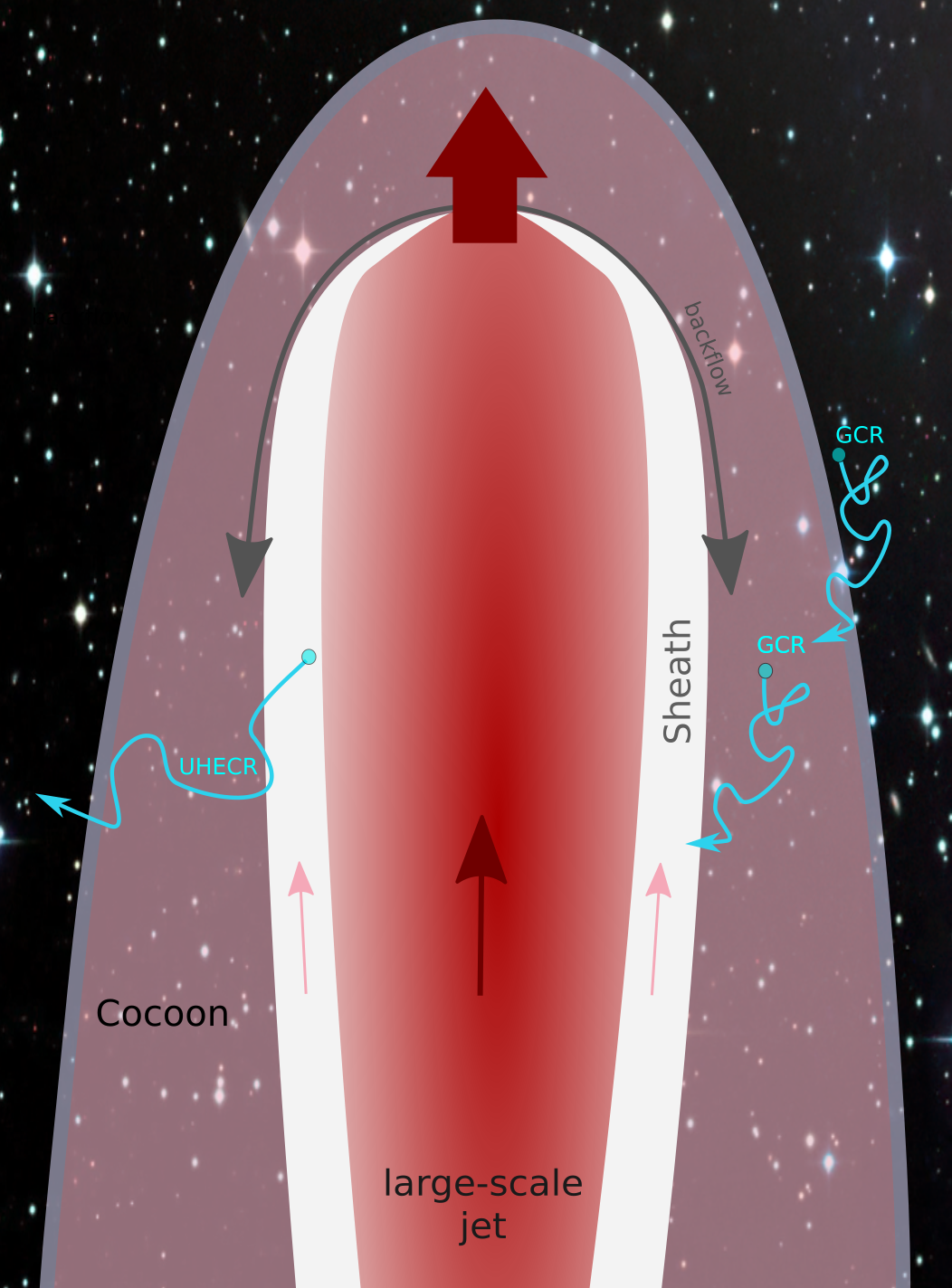}
\caption{Sketch of an UHECR model for AGN jets, in which some fraction of galactic-type cosmic 
rays (GCR) permeating the halo are swept up by the jet ("injected") and shear-accelerated to 
ultra-high energies in its jet-sheath structure. Stochastic (2nd order Fermi) acceleration by turbulence 
in the cocoon and first-order Fermi acceleration by shocks in back-flowing regions may further add to 
particle energisation.}
\label{AGN_zoom} 
\end{center}
\end{figure}

Fast shear flows are conducive to efficient particle acceleration by a variety of means, see e.g. 
ref.~\cite{Rieger2019} for an introduction and review. Prominent scenarios include Fermi-type 
shear particle acceleration in non-gradual (discontinuous), e.g., refs. \cite{Ostrowski1998,
Ostrowski2000,Caprioli2015,Kimura2018,Mbarek2021}, and gradual (continuous), e.g., 
refs.\cite{Rieger2004,Rieger2016,Liu2017,Webb2018,Webb2019,Webb2020,Wang2021,
Merten2021,Rieger2019,Rieger2022}, jet velocity shear flows. Particle acceleration in this 
framework basically draws on the kinematic flow velocity difference ($\Delta u = \Delta \beta\,c$) 
that a particles experiences as it moves across the jet shear. AGN jets may be seeded with 
protons and heavier nuclei by, e.g., internal entrainment from stars intercepted by the jet, 
e.g. \cite{Bosch2012,Wykes2015}, or by the pick-up of Galactic type (SNR accelerated) 
cosmic rays permeating its halo environment, e.g. \cite{Caprioli2015,Kimura2018}, cf. 
Figure~\ref{AGN_zoom}. 
{\it Non-gradual shear particle acceleration} then relies on the fact that, provided the particle 
distribution would remain quasi-isotropic near a shear discontinuity (e.g., a flow velocity "jump"), 
the mean fractional energy change for a CR particle crossing and re-crossing the velocity shear 
(full cycle) is given by, e.g. \cite{Rieger2004,Rieger2007}
\begin{equation}
\left< \Delta E/E \right> \sim 2 \Gamma^2\beta^2\,, \label{gain}
\end{equation}
similar to what is expected for the first Fermi cycle at relativistic shocks \cite{Gallant1999}. 
This might suggest that cosmic-ray seed particles could be boosted to very high energies in a 
velocity shear that is highly relativistic ($\Gamma \gg 1$, cf. \cite{Caprioli2015,Mbarek2021}), 
while for a non-relativistic shear flow ($\Gamma \sim 1$) only modest gains $\propto \beta^2$ 
are obtained. In reality the situation is likely to be more complex: For relativistic flow velocities 
($\beta \simeq 1$), the emerging anisotropy of the particle distribution has to be appropriately 
modelled and taken into account. For repeated shear crossings, the principal effects of this is 
a reduction in efficiency, see e.g. refs.~\cite{Ostrowski1998,Ostrowski2000}. 
Formally, this corresponds to a multiplication of the rhs of eq.~(\ref{gain}) by an efficiency factor 
$\eta_s$ significantly below unity.\\ 
Knowing the average energy change, the characteristic acceleration timescale can then be 
approximated by
\begin{equation}
t_{\rm acc} \simeq \tau / \left<\Delta E/E \right>\,, \label{tacc}
\end{equation}
where $\tau$ here denotes the mean cycle time (into and out of the shear).
Note that, depending on the considered set-up, e.g., for non-gradual shear at a jet side boundary, 
$\tau$ might actually be dominated by the turbulence properties in the cocoon, i.e. by the 
(diffusion) time a CR particle needs to return to the jet boundary shear \cite{Kimura2018}. If 
the transition layer becomes larger, particle scattering will occur within the shear layer, facilitating 
{\it gradual shear particle acceleration} within it, see e.g. refs.~\cite{Rieger2004,Rieger2016,
Rieger2019,Webb2018,Webb2019,Webb2020,Wang2021}. On larger scales, this may generically 
be the case as shear flows are generally prone to shear-driven Kelvin-Helmholtz type instabilities. 
Simple stability considerations then suggest that to account for the observed structures, the 
shearing layers in large-scale AGN jets should encompass a sizeable fraction ($>10\%$) of 
the jet radius \cite{Rieger2021}.
In a microscopic picture \cite{Rieger2006}, gradual shear acceleration can be viewed as a
stochastic, second-order ($\Delta E \propto u_{\rm sc}^2$) Fermi-type particle acceleration 
process, in which the common scattering center speed, $u_{\rm sc}$, is replaced by an 
effective velocity, $\bar{u}$, determined by the shear flow profile, i.e., by the characteristic 
flow velocity change sampled over the mean free path $\lambda$ of the particle. In the 
case of a continuous (non-relativistic) shear flow $\vec{u} =  u_z(r) \vec{e}_z$ for example, 
one has $\bar{u} = (\partial u_z/\partial r)\, \lambda$, see also ref.~\cite{Rieger2019} for a 
discussion. This implies a characteristic, fractional energy change scaling as 
\begin{equation}
\left< \Delta E/E \right> \propto \left(\frac{\bar{u}}{c}\right)^2 
                                     \propto \left( \frac{\partial u_z}{\partial r} \right)^2 
                                     \lambda^2\,.
\end{equation} 
If we again write the characteristic acceleration timescale as $t_{\rm acc} \simeq \tau / 
\left<\Delta E/E \right>$, then $t_{\rm acc}  \propto 1/\lambda$, i.e., in contrast to classical 
first-order Fermi (shock) or non-gradual shear acceleration, the acceleration timescale for 
gradual shear particle acceleration is inversely proportional to the particle mean free path 
$\lambda = c\tau$. This seemingly unusual behaviour relates to the fact that as a particle 
increases its energy ($E \simeq p c $), its mean free path increase as well (typically, 
$\lambda(p) \propto p^{\alpha}, \alpha >0$), so that a higher effective velocity $\bar{u}$ is 
experienced.\\
To illustrate the potential of fast shearing flows for UHECR acceleration, three selected
applications are highlighted in the following:
\begin{itemize}
\item {\bf Non-gradual shear particle acceleration in trans-relativistic FR~I type jets:}
Kimura et al.~\cite{Kimura2018} have reproduced the observed UHECR spectrum in a 
scenario that considers non-gradual shear particle acceleration in large-scale FR~I jets, 
see Figure~\ref{Kimura}. The reference model assumes that Galactic-type cosmic rays 
(with, e.g., CR proton densities comparable to our Galaxy) are picked up, re-cycled and 
re-accelerated to UHE energies in a mildly relativistic jet ($\beta_j = 0.7$, $B_j =300
\mu$G) surrounded by a narrow shear layer ($\Delta r/r_j \sim 10^{-2}$). Their Monte 
Carlo simulations suggest that UHECRs escaping the source, possess a rather hard 
energy spectrum ($dN/dE \propto E^{-a}, a \lppr 1$).
The chemical compositions at UHE energies (relative abundance ratio) is somewhat 
more complex, as different particle species ({\it{i}}) have different injection thresholds into the 
acceleration process ($E_{\rm inj,i} =15\,Z_{\rm i}$ TeV). Achievable maximum energy, 
$E_{\rm i,max}\simeq 1.6 Z_i$ EeV, are limited by the jet size ($t_{\rm acc} \sim r_j/c$) 
and (via $\tau$ in $t_{\rm acc}$, cf. eq.~[\ref{tacc}]) dominated by the diffusion process
(assumed turbulence properties) in the cocoon. These numbers are in principle 
compatible with requirements on source energetics for FR~I type sources, cf. also
eq.~(\ref{lower_limit}). Given an FR~I number density of $n_{\rm FRI} \sim10^{-5}-10^{-4}$ 
Mpc$^{-3}$ (cf. footnote~2) for example, an average source luminosity $L_{\rm FRI} \sim 
2\times 10^{40}-2\times 10^{41}$ erg/sec would be needed to account for the observed 
UHECR luminosity $l_{\rm UHECR} \sim 6\times 10^{43}$ erg/(Mpc$^3$ yr) \cite{Murase2019}.
Moreover, possible anisotropy constraints are relaxed, partly due to suitable UHECR 
spectral shapes, as well as a high source number density (multi-FR~I source contributions) 
with heavy composition. 
\begin{figure}[htb]
\begin{center}
\includegraphics[width=0.60 \textwidth]{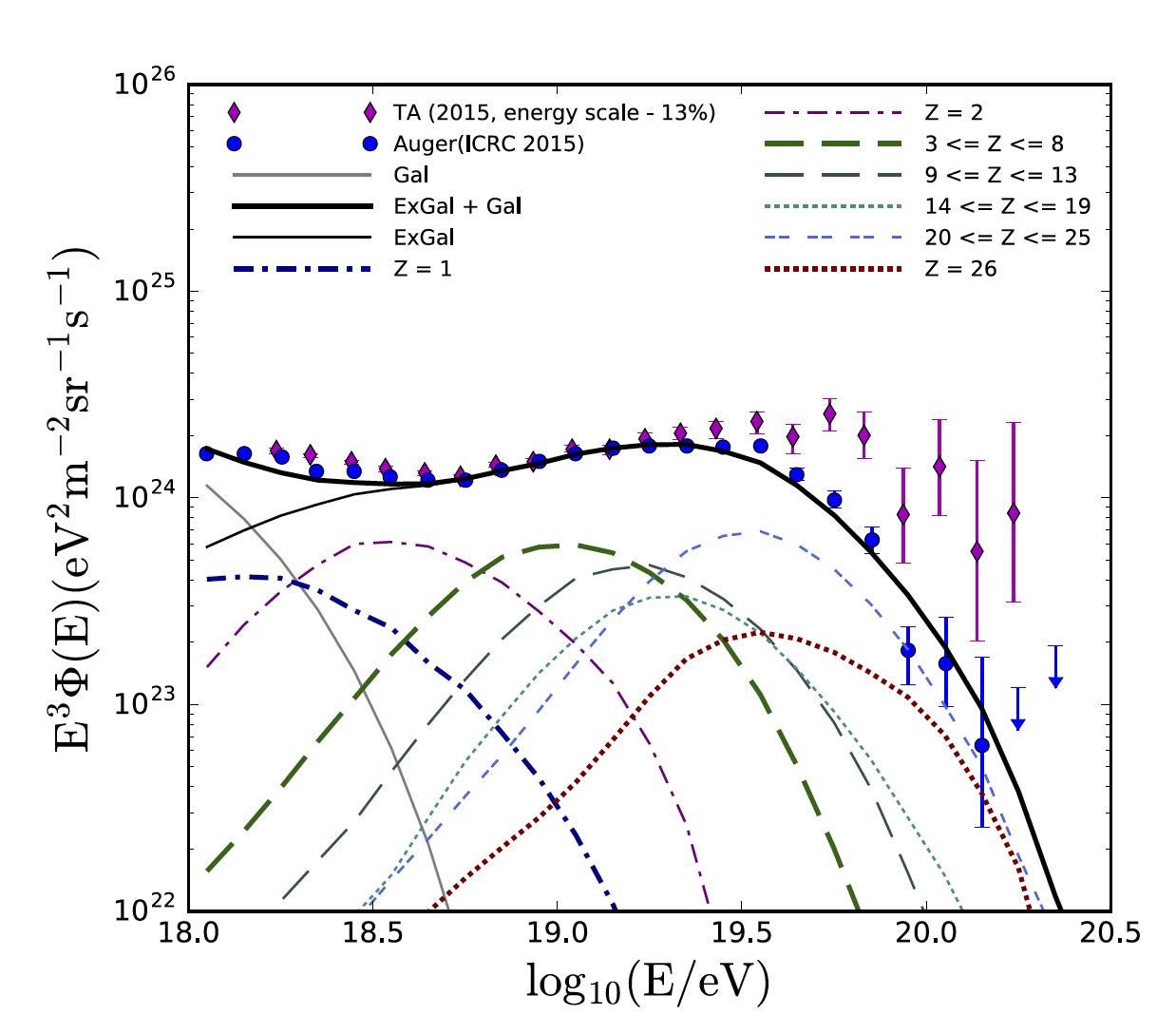}
\caption{Reconstruction of the observed UHECR spectrum in a multi-FR~I source 
scenario, assuming as reference a mildly relativistic ($\beta \simeq 0.7$) jet of width 
$r_j=0.5$ kpc surrounded by a narrow transition layer of width $\Delta r=5$ pc ($B_j
=0.3$ mG). The chemical composition at the highest energies is dominated by 
intermediate and heavy nuclei. From ref.~\cite{Kimura2018}.}
\label{Kimura}
\end{center}
\end{figure}
The results shown are, however, sensitive to the chosen cocoon properties (i.e., the 
assumed cocoon size and turbulence scale) and, in particular, dependent on a narrow 
velocity transition layer $\Delta r \sim r_j/100$ (defining the energy threshold for CR 
injection). Enlarging $\Delta r$ to satisfy jet stability constraints \cite{Rieger2021}, for 
example, is likely to affect the outcome. Nevertheless, these simulations illustrate 
that non-gradual shear acceleration in an ensemble of FR~I's (and not only FR~II's) 
could in principle play an important role in UHECR acceleration.\\ 

\item {\bf Non-gradual shear particle acceleration in large-scale FR~II type jets:}
Powerful AGN jets can remain significantly relativistic ($\Gamma\gg1$) on large 
scales, allowing for a substantial particle energy change per cycle if strong flow
gradients are experienced over the mean free path of a particle, cf. eq.~(\ref{gain}). 
In fact, given suitable conditions, significant $\Gamma^2$-type boosts are 
possible across quasi-narrow shears, in which case only a few cycles ("shots") 
would be needed to boost Galactic-type cosmic rays to UHE energies (sometimes 
referred to as "espresso" acceleration, e.g., ref.~\cite{Caprioli2015}). 
In this context, Mbarek \& Caprioli \cite{Mbarek2019,Mbarek2021} have recently 
studied the recycling of energetic cosmic rays in a relativistic ($\Gamma\simeq7$) 
large-scale jet by following their test particle trajectories in a simulated (PLUTO) 
3d-MHD jet environment, assuming a homogeneous ambient medium setup, 
see e.g., Figure~\ref{Mbarek}.
\begin{figure}[htb]
\begin{center}
\includegraphics[width=0.52 \textwidth]{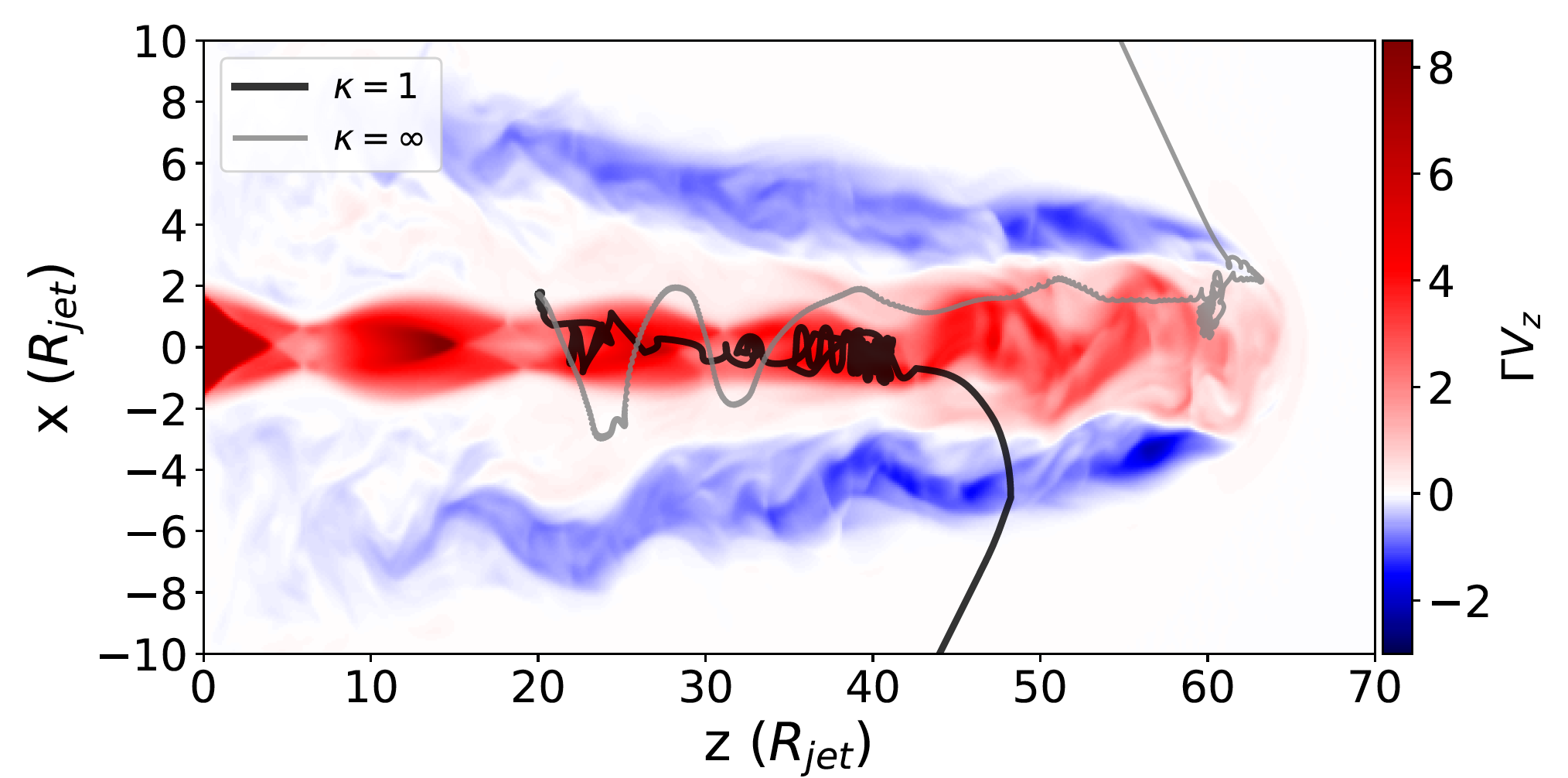}
\includegraphics[width=0.47 \textwidth]{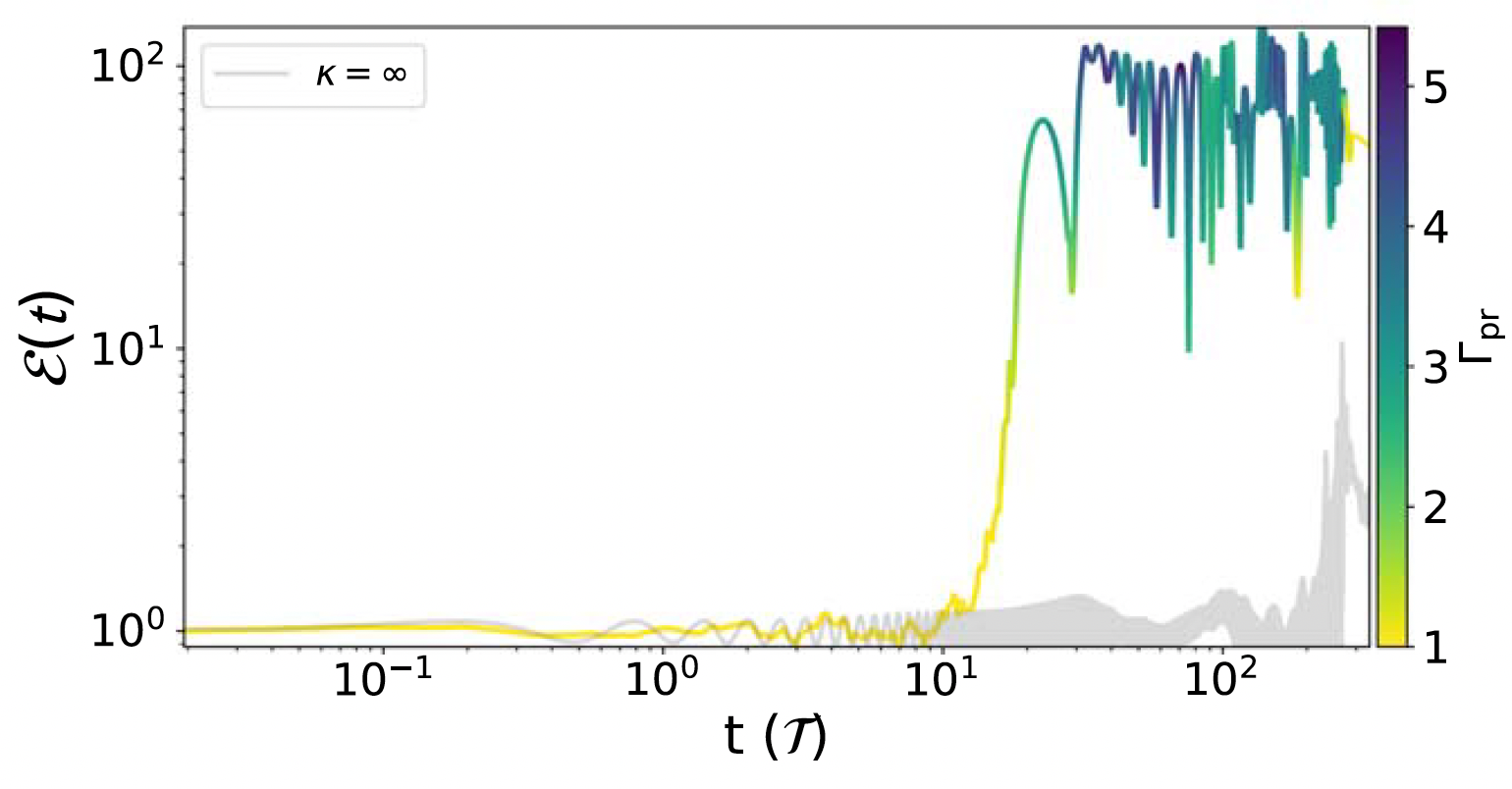}
\caption{{\it Left:} Two-dimensional projection of exemplary trajectories of energetic
particles, injected with initial Larmor radius $r_L\sim r_j/14$ into the simulated MHD 
jet environment. {\it Right:} Evolution of the relative energy gain of these particles, 
$\epsilon,$ in terms of their initial relativistic gyro-period $\tau$, colour-coded with 
the Lorentz factor $\Gamma_{\rm pr}$ probed. In 3d, the particle following the grey 
trajectory, only experiences, in the absence of scattering ($\kappa=\infty$), 
some moderate velocity differences (hence, gains) towards the end, and eventually 
becomes unconfined (escapes) as the initial jet magnetic field is assumed to 
decrease, approximately as $B(z) \propto 1/z$ along the jet, and a toroidal field 
scaling $B(r\geq r_j)\propto 1/r$ and $B(r \leq r_j)\propto r$ has been employed. 
The particle following the black trajectory undergoes strong spatial (Bohm) 
diffusion ($\kappa=1$), experiences the fast jet spine and significantly gains 
energy (by up to a factor of $\sim100$) in these interactions. From 
ref.~\cite{Mbarek2021}.}
\label{Mbarek}
\end{center}
\end{figure}
Given the fact that MHD jet simulations generally come with limited resolution 
(i.e., here of the order $\sim r_j/10$, where $r_j$ is the jet radius), a suitable 
(sub-grid) prescription needs to be designed to explore the effect of unresolved 
turbulence on the cosmic-ray transport. 
In the considered example, a gyro-dependent spatial diffusion coefficient $D = 
\kappa r_L c/3$, with $r_L \propto \gamma$ the Larmor radius, has been 
employed (with momentum-diffusion being neglected), exploring cases where 
$\kappa=1,10,100,1000$, with $\kappa=1$ corresponding to Bohm diffusion, 
and $\kappa \rightarrow \infty$ implying no diffusion/scattering.
The results show that incorporation of particle scattering (spatial diffusion) 
generally allows more particles to reach higher energies by probing the jet 
spine, making acceleration more efficient and resulting in a hardening of 
the injection spectrum. These features appear akin to what is expected in 
gradual- and non-gradual shear particle acceleration. To adequately assess
the implications, an extension of the simulation framework to more general
conditions will be helpful. As of now, comparison appears limited to, e.g., 
gyro-dependent pitch-angle diffusion, and rather high-power ($L_j \gppr 
10^{44}$ erg/s), significantly magnetized ($\sigma \sim 0.6$) FR~II type 
jets. 
Nevertheless, these results already illustrate that for sufficiently large 
mean free paths multiple energy-boosts can occur in the environment of
relativistic jets, allowing to push particle energies beyond what, e.g., 
stochastic or gradual shear acceleration alone might achieve.\\

\item {\bf Gradual shear particle acceleration in relativistic AGN jets:} 
On large scales, the shearing layers around relativistic jets are likely to 
encompass a sizeable fraction of the jet radius, e.g. \cite{Rieger2021}.
Acceleration of electrons in such shearing flows has been proposed to 
account for the extended X-ray synchrotron emission that requires the 
maintenance of ultra-relativistic electrons ($\gamma_e \sim 10^8$) on 
kpc-scales, e.g., refs.~\cite{Liu2017,Wang2021}. In order to compete
with diffusive escape, efficient shear acceleration typically requires 
sufficiently relativistic flow speeds, i.e., bulk Lorentz factors of some few 
\cite{Webb2018,Webb2019,Rieger2019}. At mildly relativistic speeds,
intrinsic particle spectra become rather soft and sensitive to the underlying 
shear flow profile \cite{Rieger2022}. It seems natural to suppose that 
the underlying shear acceleration mechanism can be operative not only 
on electrons but also on cosmic rays. In principle, by modelling their 
multi-wavelength large-scale jet emission, improved constraints on the 
parameter space of individual sources can be obtained, thereby allowing 
a more precise classification \cite{Wang2021}.
Still, for a first orientation, a Hillas type approach (Sec.~\ref{constraints})
may be chosen, requiring the properties of a potential UHECR source to
be such as (i) to allow UHECRs to be confined within the shear, (ii) to  
enable particle acceleration to overcome radiative cooling (e.g., $t_{\rm 
acc} <t_{\rm syn}$) and (iii) to operate within the lifetime ($t_{\rm dyn}$) 
of the system. Figure~\ref{Rieger2} shows an example, adjusted to 
(mildly relativistic) large-scale AGN jets, assuming a linearly decreasing 
flow profile with $\Gamma_b=3$ on the jet axis. A shear-width to jet length 
ratio $\Delta r/d=0.02$ has been employed.
\begin{figure}[hbt]
\begin{center}
\includegraphics[width=0.70 \textwidth]{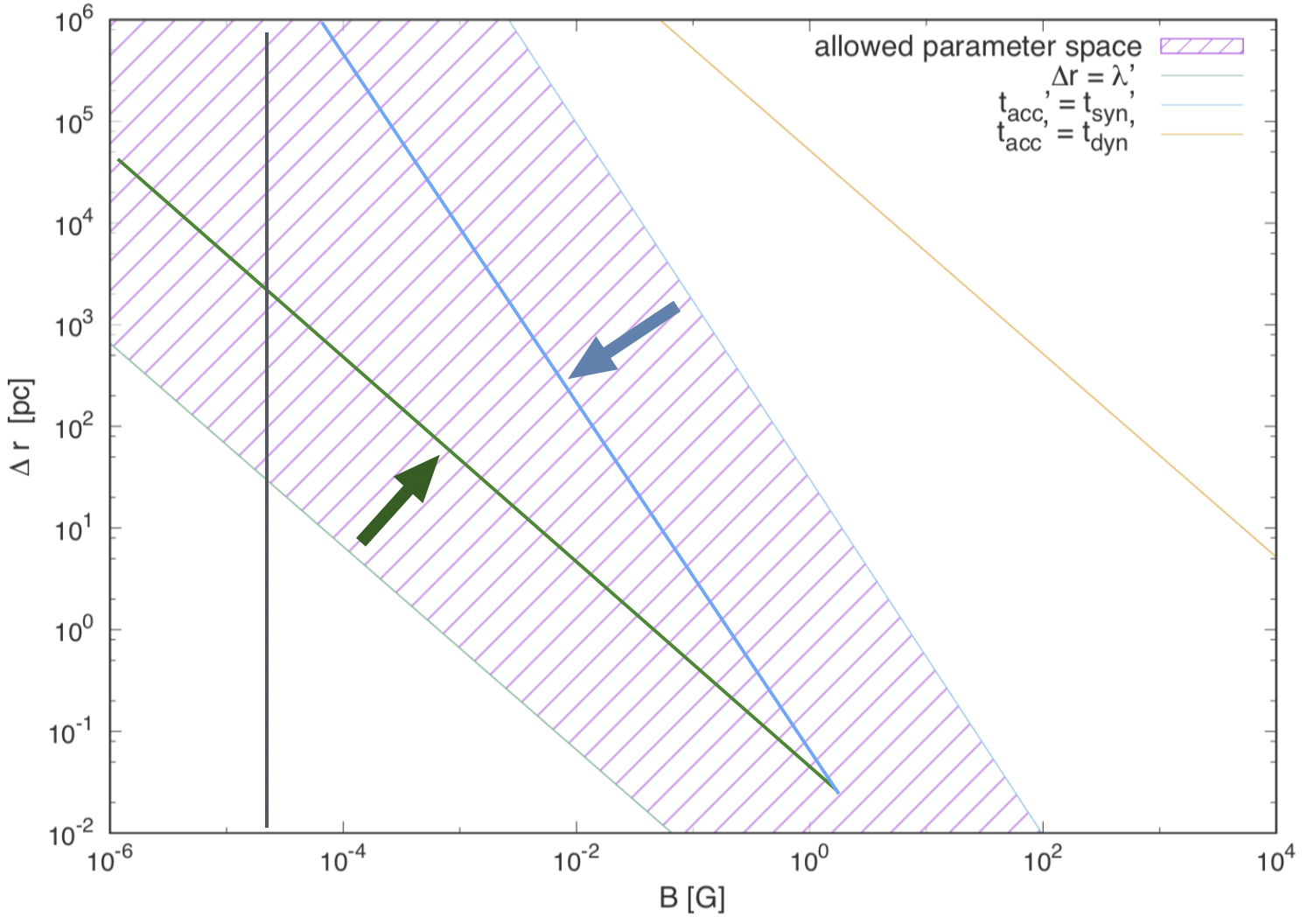}
\caption{Allowed (shaded) parameter range, i.e. magnetic field strength $B$, 
shear layer width $\Delta r$, to facilitate gradual shear acceleration of protons 
in (mildly relativistic) large-scale AGN jets, and to comply with confinement 
and synchrotron loss constraints. 
A Kolmogorov-type scaling for the particle mean free path ($\lambda \propto 
\gamma^{1/3}$) has been assumed. While for protons, EeV energies is well 
achievable, shear acceleration of protons to $\sim100$ EeV in mildly 
relativistic jet flows appears quite restricted (to the parameter space confined 
by the arrows). Following ref.~\cite{Rieger2019}.}
\label{Rieger2}
\end{center}
\end{figure}
Figure~\ref{Rieger2} suggests that shear acceleration of protons to energies 
$\sim10^{18}$ eV can in principle be achieved for a quite plausible range of 
parameters, cf. also refs.~ \cite{Liu2017,Webb2018,Webb2019,Wang2021}
The vertical (black) line, for example, denotes the regime for a Cen~A type 
source ($B\sim 20\mu$G, cf. \cite{HESS_CenA2020}), where the presence of 
a shear width of several tens of parsec would be sufficient. However, proton 
acceleration to $\sim 10^{20}$ eV in mildly relativistic shearing flows would 
seem to be less likely (requiring large, extended shears and higher magnetic 
fields). Since higher energies are possible either, in faster flows or for heavier 
particles, this would then suggest that current Cen~A type (FR~I) sources 
are contributors of heavy UHECRs, while $\sim100$ EeV protons would have 
to be associated with an earlier, more powerful (faster jet, FR~II type) stage. 
This could possibly be linked to an evolutionary scenario where during the 
lifetime of a radio galaxy (at least for some of the nearby), basic 
characteristics could change FR class from FR~II to FR~I. 
In a simple leaky-box approach for gradual shear acceleration, UHECRs 
escaping their source, can have a relatively hard spectrum, approaching 
$N(E) \propto E^{-1}$ \citep{Rieger2019}. The spectrum is expected to be 
more complex, though, if non-gradual shear acceleration becomes operative 
toward higher energies. 
\end{itemize}

\subsection{Shock acceleration of UHECRs in AGN jets} 
\begin{figure}[hbt]
\begin{center}
\includegraphics[width=0.40 \textwidth]{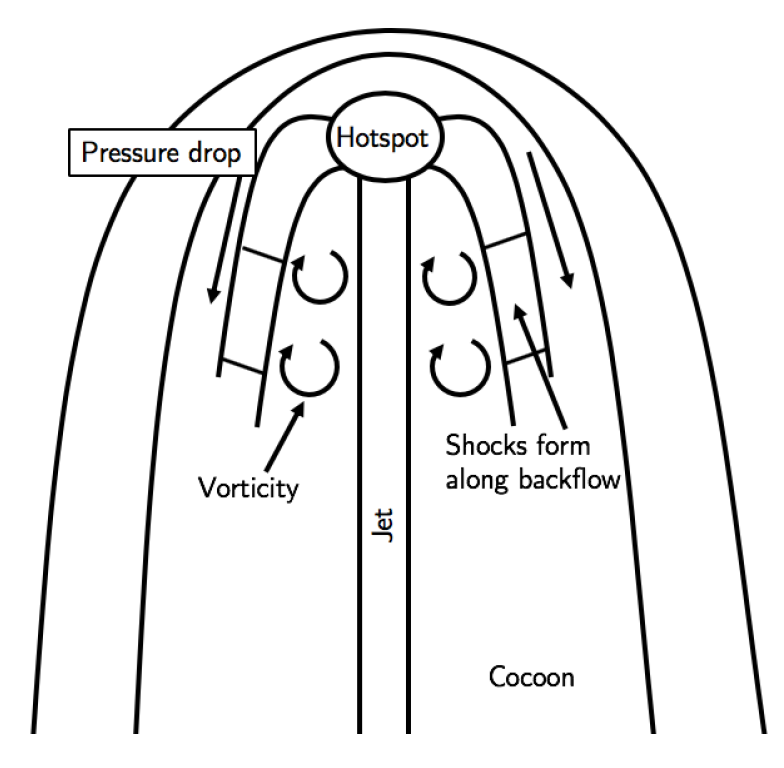}
\caption{Sketch of the considered scenario, where multiple (mildly relativistic) 
shocks in fast back-flows surrounding the large-scale jets of radio galaxies are 
considered to provide sites for efficient UHECR acceleration. From 
ref.~\cite{Matthews2019b}.}
\label{Matthews}
\end{center}
\end{figure}

Following the general considerations above (e.g., eq.~[\ref{Hillas_2}]), relativistic 
speeds are seemingly conducive for efficient UHECR acceleration. This, however, 
does not necessarily apply to diffusive shock acceleration. 
In fact, in many circumstances, particle acceleration at highly relativistic shocks has 
turned out to be problematic, see e.g. refs~\cite{Lemoine2010,Sironi2011,Bell2018,
Vanth2020} for review. This is partly due to the fact that ultra-relativistic (magnetized)
shocks are generically perpendicular (i.e., with downstream magnetic field quasi 
perpendicular to the shock normal, thus preventing particles from diffusing back 
upstream) and that the isotropization of particles ahead of the shock is no longer 
guaranteed (i.e., there is not enough time to growth turbulence on scales $r_{\rm L,
UHECR}$). The situation is much more relaxed for mildly relativistic shock speeds, 
and this has led to the proposal that UHECR acceleration preferentially occurs at 
multiple (mildly relativistic) shocks in the back-flow regions surrounding the large-scale 
jets of radio galaxies, instead of at jet termination shocks \cite{Matthews2019}. 
Diffusive particle acceleration at trans-relativistic shocks is accompanied by 
an acceleration timescale approximately scaling as $t_{\rm acc} \sim 10 
\frac{\kappa_s}{u_s^2}$, where $\kappa_s$ is the spatial diffusion coefficient.
For typical numbers, the shock would have propagated a distance of up to a 
few kilo-parsecs if protons would be accelerated to $\sim100$ EeV.
Multiple shocks can in principle provide multiple opportunities for efficient particle 
acceleration, and also produce particle spectra $N(E)$ harder (flatter) than the 
canonical (single-shock) power-law with index $s\sim2-2.3$, e.g. \cite{Pope1994,
Gieseler2000}. The latter introduces some flexibility to account for the observed 
UHECR spectrum as suggested by some models \cite{PAO2017_comp}.\\
The multiple shock-model has been motivated by 2D- and 3D-hydrodynamical 
simulations of light (density contrast $\rho_j/\rho_0 \sim 10^{-5}-10^{-4}$), 
powerful ($\sim 10^{45}$ erg/sec; more akin to FR~II) jets in cluster environments, 
where significant back-flows have been found (e.g., see also \cite{Perucho2007,
Perucho2019}). Along these, multiple, moderately strong shocks have been seen 
to form, cf. Figure~\ref{Matthews} for illustration. 
The analysis of Lagrangian tracer particles in ref.~\cite{Matthews2019} shows 
that about $10\%$ of the particles pass through a shock of Mach number $M>3$, 
and $\sim5\%$ pass through multiple shocks. Typical inferred shock speeds 
are of the order of $u_s\sim0.2$ c, while estimated sizes are $R \sim2$ kpc. 
Since in this setup, particles can only escape sidewise by diffusing across the 
magnetic fields, cosmic rays could reach the Hillas energy limit in a long-lived 
back-flow, even if Bohm diffusion ($\kappa_s \simeq r_L c/3$) would not be 
achieved \cite{Bell2019}. Assuming a reference magnetic field strength of 
$B\sim 100\,\mu$G, for example, maximum cosmic-ray energies of $E_{\rm max} 
\simeq 4 \times 10^{19} Z \,(B/10^{-4} \mathrm{G})~(R/2\mathrm{kpc})$ eV, see 
eq.~(\ref{Hillas_2}), would seem achievable. The rather large magnetic field strength
employed here, could possibly be generated by magnetic field amplification as the 
plasma passes through the termination hot-spot into the back-flow 
\citep{Araudo2018}. While the general framework is clear, full high-resolution MHD 
simulations might be needed to validate numbers (e.g., field strengths, shock size, 
compression ratio) given the fact that the magnetic energy contribution would 
no longer seem to be negligible, and that weak shocks would be rather poor 
accelerators. Be it as it may, the multiple shock-scenario provides a stimulating 
rationale for the relevance of intermediate powerful, nearby radio galaxies in 
shaping the observed UHECR spectrum, see, e.g., ref.~\cite{Matthews2018}.

\section{Conclusion and Perspectives}
Current anisotropy results indicate that both, starburst galaxies and jetted AGNs could 
be (or host) the long-sought sources of UHECRs. While theoretical considerations suggest 
that starburst winds are less promising sites (by being not powerful enough), radio galaxies 
appear in principle capable of satisfying the relevant (Hillas-Type) requirements. 
In the AGN context, a variety of possible (not mutually exclusive) UHECR acceleration 
mechanisms have been explored in recent times. Among them, Fermi-type particle 
acceleration at trans-relativistic (internal) shocks and/or in shearing, relativistic jet flows 
appear most promising. In general, nearby FR~I type radio galaxies (or restarted jets in 
close-by galaxies) could provide suitable environments in this regard. 
From a theoretical (particle acceleration) and phenomenological (UHECR anisotropy)
point of view, insights from multi-messenger astrophysics \cite{Meszaros2019} become 
important to obtain a more detailed picture, and to adequately assess the potential of specific 
source candidates. If the UHECR composition would become heavier (with, e.g., $Z\geq 
10$) towards highest energies, as indicated by the recent Pierre Auger results, nearby 
FR~I's, such as Cen~A, M87 and Fornax A, could probably meet the minimum power 
requirement already in activity phases comparable to the current ones. Since UHECRs are 
likely to escape rather slowly from their source environment, a significant proton contribution 
(if any) might possibly be related to a higher source stage in the past. There are indications, 
for example, that the giant lobe evolution in Cen~A has been shaped by an enhanced activity 
in the past, making these lobes a potential reservoir of UHECR protons.\\ 
Further experimental constraints on the chemical composition and anisotropies of UHECRs, 
along with physically-motivated propagation studies will be important to eventually conclude 
about the astrophysical sources of UHECRs.\\

\noindent
{\small Acknowledgement: I am grateful for inspiring discussion on UHECRs with 
Denis Allard and Sergey Ostapchenko during the 18th Rencontres du Vietnam 
Workshop. I have much benefited from insights provided by Felix Aharonian and 
Tony Bell. I also grateful to thank Andrew Taylor and Jonathan Biteau for helpful 
comments on the manuscript. I would like to thank Kumiko Kotera, Caina de 
Oliveira, James Matthews, Rostom Mbarek, Khota Murase and Sjoert van Velzen 
for allowance to use figures from their papers. Funding by a DFG Fellowship 
RI 1187/8-1 is gratefully acknowledged.}

\bibliography{paper_biblio.bib}

\end{document}